\documentclass[%
 reprint,showpacs,
 amsmath,amssymb,
 aps,nofootinbib,superscriptaddress,
numbers
]{revtex4}
 \usepackage{ulem}
    \usepackage{float}
    \usepackage{natbib}
\usepackage{graphicx}
\usepackage{dcolumn}
\usepackage[colorlinks=true, linkcolor = red, citecolor = blue]{hyperref}
\usepackage{cancel}
\usepackage{subcaption}
\usepackage{float}
\usepackage{xcolor}

\begin{document}
\newcommand{\lu}[1]{#1}
\newcommand{\Quita}[1]{\textcolor{lightgray}{#1}}
\newcommand{\dario}[1]{\textcolor{blue}{#1}}
\newcommand{\JCH}[1]{\textcolor{teal}{#1}}
\newcommand{\quitar}[1]{\textcolor{purple}{\cancel{#1}}}
\title{Long-wavelength  nonlinear perturbations of a complex scalar field}

\author{Luis E. Padilla}
\email{lepadilla@icf.unam.mx}
\affiliation{Instituto de Ciencias F\'isicas, Universidad Nacional Aut\'onoma de M\'exico,
Apartado Postal 48-3, 62251 Cuernavaca, Morelos, M\'exico.}
\affiliation{Mesoamerican Centre for Theoretical Physics,
Universidad Aut\'onoma de Chiapas, Carretera Zapata Km. 4, Real
del Bosque (Ter\'an), Tuxtla Guti\'errez 29040, Chiapas, M\'exico}

  \author{Juan Carlos Hidalgo}
  \email{hidalgo@icf.unam.mx}
\affiliation{Instituto de Ciencias F\'isicas, Universidad Nacional Aut\'onoma de M\'exico,
Apdo. Postal 48-3, 62251 Cuernavaca, Morelos, M\'exico.}  \author{Darío Nú\~nez}
  \email{nunez@nucleares.unam.mx}
\affiliation{Instituto de Ciencias Nucleares, Universidad Nacional Aut\'onoma de M\'exico,
Circuito Exterior C.U.,  M\'exico City 04510, Mexico}
\date{\today}

\begin{abstract}

We study the evolution of nonlinear superhorizon perturbations in a universe dominated by a complex scalar field. The analysis is performed adopting the gradient expansion approach, in the constant mean curvature slicing. We derive general {solutions} valid to second order in the ratio $H^{-1}/L$ for scalar field inhomogeneities of size $L$ subject to an arbitrary canonical potential. We work out explicit solutions for the quadratic and the quartic potentials, and discuss their relevance in setting initial conditions required for the simulations of  primordial black hole formation. 
\end{abstract}

\pacs{PACS}
\maketitle

\section{Introduction}

For decades, cosmologists have been interested in scalar fields (SFs) and their role in the evolution of our Universe \citep{zee2010quantum,dodelson2003modern}. The dynamics of such fields is usually described by the Einstein-Klein-Gordon (EKG) system of equations, which can be regarded as the relativistic generalization of Schrödinger-Poisson (SP) or Gross-Pitaesvkii-Poisson (GPP) systems (the second case arising when a self-interaction between particles is considered). The system was initially studied in the context of boson stars \citep{bs1,bs2,bs3,bs4,bs5,bs6,bs7,bs9,bs10,bs11,bs12,bs13,bs14}, {inferred from the axion field} -- a pseudo-Nambu-Goldstone boson of the Peccei-Quinn phase transition --, which was originally proposed to solve the strong CP problem in QCD. Such SFs have shown a variety of dynamical properties in a cosmological context; proving useful as important components of the Universe, such as: dark matter (scalar field Dark Matter, "SFDM", see Refs. \citep{Matos:2000ng,Matos:1998vk,Matos:1999et,Peebles2000,Goodman:2000tg,Sahni:1999qe,charge2,Arbey:2003sj,Cedeno:2017sou}, and also Refs. \citep{rev1,rev2,Marsh:2015xka,rev4,niemeyer2019small,RS,ferreira2020ultra} for comprehensive reviews of this model), a variety of models acting as Dark Energy (DE) \citep{de1,de2,de3}, such as quintessence \citep{quintessence,Zlatev:1998tr,Corasaniti:2002vg, de4}, or phantom DE \citep{de5,de6,de7,de8}. Tachyonic instabilities arise ubiquitously in SFs models \citep{de9,de10,de11} and, most successfully, the majority of inflationary models  \citep{Linde1982,inf3,inf4,Guth,Lucchin:1984yf,Liddle:1999mq,Ratra:1989uz,DiMarco:2018bnw,inf2} {(see also Refs. \citep{Vazquez:2018qdg} for a comprehensive review)}, among other possible realizations as components of the matter sector. 

In the context of dark matter it is usually assumed that the SFDM could be a real or complex SF minimally coupled to gravity. In the simplest scenario,  the SFDM is a real scalar field (RSF) subject to a quadratic or mass potential. It has been demonstrated that the model is able to reproduce all the predictions of the standard cosmological model -- the so-called $\Lambda$-\textit{cold dark matter} ($\Lambda$CDM) model -- with the advantage of solving some of the problems  at small scales that $\Lambda$CDM implies; namely, the overproduction of satellite dwarf galaxies within the local group \citep{16,9,18} and the \textit{cusp-core} problem \citep{matos2007flat,robles2012flat}. 
%
%
%
Additional phenomenology appears when modifications of the simplest model are considered. For example, the complex scalar field (CSF) presents a cosmological evolution at early times that departs from that of a real field \citep{li2014cosmological}. Moreover, its perturbations present a wider instability band than its real field counterpart \citep{Carrion:2021yeh}. Despite all the success of SFDM, formal solutions for large scales (super-Hubble modes) are scarce and largely required for initial conditions in simulations of structure formation. This is a strong motivation for the present work. 

A phase analogous to a Universe dominated by SFDM arises in the primordial universe, at much higher energy scales, in the reheating period -- the transition period from inflation to the standard hot big bang cosmology \citep{reh1,reh2,reh3,reh4,Kofman:1997yn}. A few reheating models propose a stage where one or more scalar fields oscillate around the bottom of their potential. Just as in SFDM, the universe dominated by a fast-oscillation field evolves effectively as a dust-dominated space. In such a scenario, primordial compact structures may form, analogous to those present in the SFDM model, but at much earlier times. With this in mind, Ref. \citep{sfdmrh1} \citep[see also Refs. ][]{sfdmrh2,Eggemeier:2020zeg} argues that this ``primordial structure formation" process could be completely analogous to the structure formation in SFDM, since from a computational point of view the only difference between both processes is given by the region in the parameter space of the  model (and one may argue that the matter power spectrum differs in each scenario). In this way, several of the results of SFDM have been adopted for reheating models, showing that the reheating process could have taken place in a universe filled with inhomogeneities as a result of the fragmentation of the inflaton and formation of inflaton clusters.  

A strong motivation for the present work is the possibility of structure formation during the reheating period, as well as the analogous process in the SFDM model. The inhomogeneities seeding the structure right after primordial inflation start their evolution on scales above the cosmological horizon (super-Hubble modes). Thus, the complete dynamics cannot be captured through a Newtonian system. In order to provide adequate solutions from the more general EKG system, it proves useful to adopt a gradient expansion of the evolution equations (although higher order perturbative KG equations are available \cite{Malik:2006ir}). {The gradient expansion formalism features an advantage over the standard picture (the well known linear theory of cosmological perturbations): the fact that the former is not restricted to configurations with small amplitudes (linear). Instead, the only requirement of the gradient expansion is for inhomogeneities to be larger than the cosmological horizon. This description is, thus, useful to model any kind of superhorizon perturbations, including those which enter the cosmological horizon with a large amplitude, typical of configurations which form primordial black holes (PBHs).} 

In this paper we study the initial evolution stages of CSF inhomogeneities as solutions of the EKG system, which serve as initial conditions for structure formation at, say, the reheating period. While historically the inflaton has been assumed as a RSF, several models consider instead a complex field \citep[see for example, Refs.][]{compinf1,compinf2,compinf3,compinf4,compinf5,compinf6,compinf7,compinf8,compinf9,compinf10,compinf11}. Complex fields appear naturally in extensions of the standard model of particle physics and they could have played an important role in the evolution of the Universe at high energies. Furthermore, due to the well known differences between real and complex SFDM, it is natural to explore the particularities that a CSF inflaton bring to the "primordial structure formation". 
Our study is relevant to the origin of complex scalar field structures in a cosmological background, when the evolution of inhomogeneities can be accounted for by the gradient expansion formalism \citep{salopek1990nonlinear}, in which spatial derivatives are assumed to be small compared to time derivatives. Specifically, an expansion parameter $\epsilon$ for coordinate derivatives is introduced in the EKG system of equations, which is then solved order by order in powers of this parameter. Our purpose is, thus, to formulate nonlinear solutions for a CSF which can later be used as initial conditions for general relativistic evolution codes which work with the EKG system, instead of the SP or GPP. The gradient expansion for a RSF has been previously studied in the literature in \citep{tanaka2007gradient} {(see also Refs. \citep{kodama1998evolution}, \citep{Takamizu:2018uty} for the generalization to a general kinetic term for the RSF, \lu{and \citep{Sasaki:1998ug,Malik:2005cy} for the multi-field case})}. As for the complex scalar field,  long-wavelength, but subhorizon and linear perturbations of a complex scalar field in the context of $P(X)$ theories, has been studied in \cite{Babichev:2018twg}.  We intend here to generalize the long-wavelength analysis to the case of a complex field.

The paper is organized as follows. In Sec. \ref{section2} we present the energy-momentum tensor and the evolution equations of a CSF subject to an arbitrary potential $V(|\varphi|^2)$. In that same section, we rewrite the energy-momentum tensor of the CSF as a perfect fluid, with a word of warning. Thereafter, we adopt for our calculation the perfect fluid representation of the CSF. In Sec. \ref{section3} we present the constraints and evolution equations we work with. Specifically, we present the EKG system in the so-called $3+1$ formalism to then decompose our system of equations in a cosmological conformal decomposition. In Sec. \ref{section4} we introduce the gradient expansion formalism to derive the $O(\epsilon^2)$ equations valid for a CSF. We then impose the \textit{constant mean curvature} (CMC) slicing to simplify our equations as the standard practice dictates \cite{shibata1999black, harada2015cosmological,tanaka2007gradient,tanaka2007gradient2}. In that same section, we  present the solutions of our system of equations which describe the evolution of superhorizon inhomogeneities of a CSF. In Sec. \ref{section5} we apply our findings to some simple realizations of the CSF potential, {namely a quadratic and a quartic potential,} while Sec. \ref{conclusions} is where we draw our conclusions. 

\section{Complex scalar field in a perfect fluid form}\label{section2}

Through this work, we consider a minimally coupled CSF $\varphi$ subject to a generic potential $V(|\varphi|^2)$. Its energy-momentum tensor is given by
\begin{equation}\label{emfield}
    T_{\mu\nu} = \frac{1}{2}\nabla_\mu\varphi^*\nabla_\nu\varphi+\frac{1}{2}\nabla_\nu\varphi^*\nabla_\mu\varphi-g_{\mu\nu}\left[\frac{1}{2}\nabla^\sigma\varphi^*\nabla_\sigma\varphi+V(|\varphi|^2)\right].
\end{equation}
The equation of motion for $\varphi$ is given by the KG equation:
\begin{equation}\label{kgsf}
    \nabla_\mu\nabla^\mu\varphi = 2V'(|\varphi|^2)\varphi \ \ \ \Rightarrow \ \ \ \partial_\mu\left(\sqrt{-g}\partial^\mu \varphi\right) = 2\sqrt{-g}V'(|\varphi|^2)\varphi,
\end{equation}
where we have defined $V'(|\varphi|^2)$ as
\begin{equation}\label{vp}
V'(|\varphi|^2)\equiv \frac{dV(|\varphi|^2)}{d|\varphi|^2},
\end{equation}
with $g\equiv \mathrm{Det}(g_{\mu\nu})$, and $g_{\mu\nu}$ is the 4-metric tensor. Additionally, the CSF satisfies the 4-current conservation equation 
\begin{equation}\label{4current}
    \nabla_\mu \mathcal{J}^\mu = 0, \ \ \ \ \ \ \ \ \text{where} \ \ \ \ \ \ \ \ \mathcal{J}^\mu \equiv -i\left[\varphi^*\nabla^\mu\varphi-\varphi\nabla^\mu\varphi^*\right].
\end{equation}

We can rewrite the energy-momentum tensor \eqref{emfield} in a perfect fluid form:
\begin{equation}
T_{\mu\nu} = (\rho+p)u_\mu u_\nu +pg_{\mu\nu},    
\end{equation}
{if we follow the correspondences}
\begin{subequations}
\begin{equation}\label{preassure}
    p = -\left(\frac{1}{2}\nabla^\sigma\varphi^*\nabla_\sigma\varphi+V(|\varphi|^2)\right), \ \ \ \rho = -\frac{1}{2}\nabla^\sigma\varphi^*\nabla_\sigma\varphi+V(|\varphi|^2),
\end{equation}
and with $u_\mu$ satisfying the condition
\begin{equation}\label{equij}
u_\mu u_\nu = \frac{\frac{1}{2}\nabla_\mu\varphi^*\nabla_\nu \varphi+\frac{1}{2}\nabla_\mu\varphi\nabla_\nu \varphi^*}{-\nabla^\sigma\varphi^*\nabla_\sigma\varphi}.
\end{equation}
\end{subequations}
From this last correspondence one can verify the normalization $u^\mu u_\mu = -1$.

{The hydrodynamic representation is a common practice in the study of cosmological scalar fields. It was first introduced by Madelung in Ref. \citep{madelung1927quantentheorie}. In his work, Madelung managed to rewrite the Schr\"odinger equation -- which results from applying the nonrelativistic limit to the KG equation -- in a Euler-like system of equations for an irrotational fluid with
an additional quantum potential arising from the finite value of $\hbar$. Since then, this hydrodynamic representation has been used in many contexts to study the CSF such as a cosmological component in the background \citep{charge4,complexsf4,Carvente:2020aae}, its linear perturbations \citep{18,jeans2, jeans3}, or in stationary configurations describing galaxies \citep{mipaper,chavanismipaper,RS,shapiro03}, among others. In the present work, we find it convenient to use this representation to study nonlinear superhorizon perturbations for the CSF. {We have, in fact, verified that the correspondence is valid up to third order in the gradient expansion featured here.} }
{The reader must be warned, however, that the scalar field is not exactly a fluid, and the analogy must be verified for each case. {In particular, the equation of state is an ill-defined concept for a scalar field, in that one cannot, in general, write the pressure as a linear function of the matter density, much less propose a given relation. Instead,}  one has to solve the KG equation, Eq.~(\ref{kgsf}), then construct the associated pressure and density, and only then derive a relation between them, (see \cite{Carvente:2020aae} as an example of this procedure).} {With this in mind it is clear that, despite the fact that in first instance we strongly rely on the hydrodynamical representation of the scalar field, at some point in our analysis, when seeking for solutions that govern the cosmological evolution of nonlinear cosmological perturbations for the CSF, it will be necessary to restore the equations to their form in terms of the field variables, in order to close the system (in the absence of an equation of state). The details of this procedure are presented below.}

\section{{Basic equations}}\label{section3}

\subsection{3+1 formalism}

{Before presenting the system of equations that describe a CSF in a cosmological context, we start by rewriting the EKG equations in a convenient form. That is, we shall rewrite the EKG equations in a 3 + 1 formalism, } where the space-time line element  is written in the following form:
\begin{equation}\label{metric3+1}
    ds^2 = -\alpha^2dt^2+\gamma_{ij}(dx^i+\beta^idt)(dx^j+\beta^jdt).
\end{equation}
Here $\alpha$, $\beta^i$, and $\gamma_{ij}$ are the lapse function, shift vector, and spatial metric, respectively. Latin indices range from 1 to 3 denoting space coordinates and are dropped and raised by $\gamma_{ij}$ and $\gamma^{ij}$, unless otherwise specified, whereas Greek indices range from 0 to 3 denoting space-time coordinates. The space-time metric and its inverse are given by
\begin{equation}
    g_{\mu\nu} = \begin{pmatrix}
    -\alpha^2 +\beta_i\beta^i & \beta_i \\
    \beta_j &\gamma_{ij}
    \end{pmatrix}, \ \ \ \ \ \ \ \ g^{\mu\nu} = \begin{pmatrix}
    -\frac{1}{\alpha^2}  & \frac{\beta^i}{\alpha^2} \\
    \frac{\beta^j}{\alpha^2} &\gamma^{ij}-\frac{\beta^i\beta^j}{\alpha^2}
    \end{pmatrix}.
\end{equation}
Then $g=-\alpha^2\gamma$, where $\gamma \equiv {\mathrm{Det}}(\gamma_{ij})$. The covariant and contravariant components of the normal unit vector to the $t=$const hypersurface $\Sigma$ are given by
\begin{equation}
    n_\mu = (-\alpha,0,0,0), \ \ \ \ \ \ \ \ n^\mu = \left(\frac{1}{\alpha},-\frac{\beta^i}{\alpha}\right),
\end{equation}
respectively. On the other hand, the projection tensor to $\Sigma$ is defined as $h^\mu_{\nu}\equiv \gamma^\mu_{\nu}+n^\mu n_\nu$.

In general, we can decompose the stress-energy tensor for the matter field $T_{\mu\nu}$ in the following form
\begin{equation}
    T_{\mu\nu} = En_\mu n_\nu +J_\mu n_\nu + J_\nu n_\mu +S_{\alpha\beta}h_\mu^\alpha h^\beta _\nu.
\end{equation}
where
\begin{equation}\label{energy_definitions}
    E\equiv T_{\mu\nu}n^\mu n^\nu, \ \ \ \ J_\alpha\equiv -T_{\mu\nu}h^\mu_\alpha n^\nu, \ \ \ \ S_{\alpha\beta}\equiv T_{\mu\nu}h_\alpha^\mu h_\beta^\nu.
\end{equation}
With these new definitions, the Einstein equations $G_{\mu\nu} = 8\pi T_{\mu\nu}$ can be written in the following set of equations,
\begin{itemize}
    \item the Hamiltonian constraint $G^{\mu\nu}n_\mu n_\nu = 8\pi T^{\mu\nu}n_\mu n_\nu$ and momentum constraint $G^{\mu\nu}n_\mu h_{\nu i} = 8\pi T^{\mu\nu}n_\mu h_{\nu i}$:
\begin{equation}\label{hamiltonian}
        \mathcal{R}+K^2-K_{ij}K^{ij} = 16\pi E,
\end{equation}
and
\begin{equation}\label{momentum1}
    D_j K_i^j-D_i K = 8\pi J_i, 
\end{equation}
respectively, where $D_i$ and $\mathcal{R}$ denote the covariant derivative and Ricci scalar, respectively, with respect to $\gamma_{ij}$, $K_{ij}$ is the extrinsic curvature of $\Sigma$, defined as
\begin{equation}
    K_{ij}\equiv -h_i^\mu h_j^\nu n_{\mu;\nu} = -\frac{1}{2\alpha}(\partial_t\gamma_{ij}-D_j \beta_i-D_i\beta_j),
\end{equation}
and $K\equiv \gamma^{ij}K_{ij}$. In the above expression, and for the rest of this work, a semicolon denotes covariant derivative with respect to $g_{\mu\nu}$. Observe also that the above expression can be rewritten as
\begin{equation}\label{extcurv}
    \partial_t \gamma_{ij} = -2\alpha K_{ij}+D_j\beta_i + D_i\beta_j,
\end{equation}
{which represents an evolution equation for the spatial part of the metric.}
\item the evolution equations $G^{\mu\nu}h_{\mu i}h_{\nu j} = 8\pi T^{\mu\nu}h_{\mu i}h_{\nu j}$ {for the metric variables},
\begin{equation}\label{evolution}
    \partial_t K_{ij} = \alpha (\mathcal{R}_{ij}+KK_{ij})-2\alpha K_{il}K^l_j-8\pi\alpha\left[S_{ij}+\frac{1}{2}\gamma_{ij}(E-S_l^l)\right]-D_jD_i\alpha+(D_j\beta^m)K_{mi}+(D_i\beta^m)K_{mj}+\beta^mD_mK_{ij}.
\end{equation}
\end{itemize}

{As usual, the lapse $\alpha$, and shift ${\beta}^i$, remain undetermined due to the covariant character of the theory, and one can choose them as they  better suit the specifics of the scenario under study.\footnote{{In the cosmological case, it is natural to choose a constant lapse and a zero shift vector (see \cite{Alcubierre_adm}),  whereas in a black hole scenario the $1+\log$ lapse and the Gamma driver shift are better suited for the problem (see \cite{Alcubierre:2005gh} for a discussion on this subject).}}}

Similarly, we can decompose the CSF equation of motion in a 3+1 form. To do this, we could proceed in two different ways, which would be to work directly with the field representation of the energy-momentum tensor and, therefore, rewrite the KG equation \eqref{kgsf} in this $3+1$ representation, while a second possibility would be to rely on the perfect fluid representation of the CSF. In this work, we proceed by adopting the second option. \lu{We must emphasize that the hydrodynamic representation of the CSF can not be always applied to all kind of metrics. However, as mentioned earlier, we verified that this equivalence exist in our gradient expansion approximation up to order $O(\epsilon^3)$ (see later).}

Given that our energy-momentum tensor is now written in the perfect fluid form, we can define a 3-velocity $v^i$ as $v^i = u^i/u_0$. We can express then $u^\mu$ and $u_\mu$ as
\begin{subequations}
\begin{equation}
    u^0 = [\alpha^2-(v_k+\beta_k)(v^k+\beta^k)]^{-1/2}, \ \ \ \ \ u^i = u^0v^i,
\end{equation}
\begin{equation}\label{vel}
    u_0 = -u^0[\alpha^2-\beta_k (v^k+\beta^k)], \ \ \ \ \ u_i = u^0(v_i+\beta_i).
\end{equation}
\end{subequations}
From the conservation equation $\nabla_\mu T^{\mu}_\nu = 0$, we obtain {a Euler-like} system of differential equations (the same that are valid for a perfect fluid):
\begin{subequations}
\begin{equation}\label{fluid1}
    u^\mu\partial_\mu \rho +\frac{\rho+p}{\alpha\sqrt{\gamma}}\partial_\mu\left(\alpha\sqrt{\gamma} u^\mu\right)=0,
\end{equation}
\begin{equation}\label{fluid2}
    \frac{1}{\sqrt{\gamma}}\partial_t[\sqrt{\gamma}(\rho+p)\alpha u^0 u_i]+{D}_j[(\rho+p)\alpha u^0 v^j u_i] = -\alpha \partial_i p-(\rho+p)\alpha u^0[\alpha u^0\partial_i \alpha-u_j{D}_i\beta^j].
\end{equation}
\end{subequations}
{The first of these two equations represents the conservation {law}  of mass, whereas the second one is the conservation law of momentum.} The different quantities defined in Eq. \eqref{energy_definitions} are then expressed as
\begin{subequations}\label{eqs19}
\begin{equation}\label{Ef}
    E = (\rho+p)(\alpha u^0)^2-p,
\end{equation}
\begin{equation}\label{Js}
    J_i = (\rho+p)\alpha u^0 u_i,
\end{equation}
\begin{equation}\label{Sf}
    S_{ij} = (\rho+p)u_i u_j+p\gamma_{ij}
\end{equation}
\end{subequations}
where {we have used that} $u_\mu u_\nu n^\mu n^\nu=u^\mu u^\nu n_\mu n_\nu = (\alpha u^0)^2$ in the above expression and $\rho$ and $p$ are given in terms of the field variables as shown in Eq. \eqref{preassure}.
\subsection{Cosmological conformal decomposition}

In this section, {we show how to rewrite the above equations  in a cosmological context. To this end,} we review the cosmological conformal decomposition as described in Refs. \citep{shibata1999black,harada2015cosmological}. The idea of this decomposition is to assume an asymptotically spatially flat Friedmann universe; in such a case the spatial metric $\gamma_{ij}$ is decomposed as $\gamma_{ij} = \psi^4a^2(t)\tilde\gamma_{ij}$, where $\tilde \gamma\equiv \mathrm{Det}(\tilde \gamma_{ij})$ is time independent and equal to $\eta \equiv \mathrm{Det}(\eta_{ij})$, with $\eta_{ij}$ a time-independent metric of the flat tree space. The function $a(t)$ is {a scale factor of a reference universe, which we can assume to be at our location}, {whereas $\psi^4$ encrypts deviations from the Friedmann universe}. Also, the extrinsic curvature is decomposed as
\begin{equation}
    K_{ij} = A_{ij}+\frac{\gamma_{ij}}{3}K,
\end{equation}
and then $A_{ij}$ is traceless by definition. Additionally, a new tensor $\tilde A_{ij}$ is also defined as
\begin{equation}
    A^{ij} = \psi^{-4}a^{-2}\tilde A^{ij}, \ \ \ \ \ \ \ \ A_{ij} = \psi^{4}a^{2}\tilde A_{ij}.
\end{equation}
With this new decomposition, we can rewrite $\mathcal{R}_{ij}$ as follows
\begin{equation}
    \mathcal{R}_{ij} = \mathcal{\tilde R}_{ij}+\mathcal{R}^\psi_{ij},
\end{equation}
where
\begin{equation}\nonumber
    \mathcal{R}^{\psi}_{ij}\equiv -\frac{2}{\psi}\tilde D_i \tilde D_j\psi-\frac{2}{\psi}\tilde\gamma_{ij}\tilde\triangle\psi +\frac{6}{\psi^2}\tilde D_i \psi \tilde D_j \psi-\frac{2}{\psi^2}\tilde\gamma_{ij}\tilde D_k\psi \tilde D^k\psi,
\end{equation}
\begin{equation}\nonumber
    \mathcal{\tilde R}_{ij}\equiv \frac{1}{2}[-\bar\triangle \tilde\gamma_{ij}+\bar D_j\bar D^k\tilde\gamma_{ki}+\bar D_i\bar D^k\tilde\gamma_{kj}+2\bar D_k(f^{kl}C_{lij})-2C^l_{kj}C^k_{il}],
\end{equation}
\begin{equation}\nonumber
    f^{kl}\equiv \tilde \gamma^{kl}-\eta^{kl}, \ \ \ \ \ \ 
    C^k_{ij}\equiv \frac{1}{2}\tilde\gamma^{kl}(\bar D_i\tilde\gamma_{jl}+\bar D_j\tilde\gamma_{il}-\bar D_l\tilde\gamma_{ij}).
\end{equation}
In the above expressions $\tilde D_i$ ($\bar D_i$) is the covariant derivative with respect to $\tilde \gamma_{ij}$ ($\eta_{ij}$), and $\tilde\triangle\equiv \tilde\gamma^{ij}\tilde D_i\tilde D_j$ ($\bar\triangle\equiv \eta^{ij}\bar D_i\bar D_j$). Also, indices for quantities with a \lu{tilde (hat)} are lowered and raised by $\tilde \gamma_{ij}$ and $\tilde \gamma^{ij}$ ($\eta_{ij}$ and $\eta^{ij}$), respectively. Then, we have
\begin{equation}
    \mathcal{R}^\psi \equiv \gamma^{ij}\mathcal{R}^\psi_{ij} = -\frac{8}{\psi^5a^2}\tilde\triangle\psi,
\end{equation}
\begin{equation}
    \mathcal{R}^\psi_{ij}-\frac{1}{3}\gamma_{ij}\mathcal{R}^\psi = -\frac{2}{\psi}\left[\tilde  D_i \tilde D_j \psi -\frac{1}{3}\tilde \gamma_{ij}\tilde \triangle \psi\right]+\frac{6}{\psi}\left[\tilde D_i\psi \tilde D_j \psi-\frac{1}{3}\tilde\gamma_{ij}\tilde D^k\psi \tilde D_k\psi\right].
\end{equation}
\newpage
With all these new definitions, Eqs. \eqref{hamiltonian}, \eqref{momentum1} and \eqref{evolution} are rewritten as folllows.
\begin{itemize}
    \item The Hamiltonian and momentum constraints:
    \begin{equation}\label{hamiltonian_constrant}
        \mathcal{R}_k^k-\tilde A_{ij}\tilde A^{ij}+\frac{2}{3}K^2 = 16\pi E,
    \end{equation}
    \begin{equation}
        D_j\tilde A^j_i -\frac{2}{3}D_i K = 8\pi J_i.
    \end{equation}
    With the conformal decomposition, the above equations can be transformed into the form
    \begin{equation}\label{eq28}
        \tilde \triangle \psi \equiv \frac{\mathcal{\tilde R}_k^k}{8}\psi - 2\pi \psi^5a^2 E-\frac{\psi^5 a^2}{8}\left(\tilde A_{ij}\tilde A^{ij}-\frac{2}{3}K^2\right),
    \end{equation}
    \begin{equation}\label{momentum}
        \tilde D^j(\psi^6\tilde A_{ij})-\frac{2}{3}\psi ^6\tilde D_i K = 8\pi J_i\psi^6.
    \end{equation}

\item The evolution equations {for the metric variables}:
\begin{eqnarray}\label{eqA_ij}
    (\partial_t - \mathcal{L}_\beta)\tilde A_{ij} =&& \frac{1}{a^2\psi^4}\left[\alpha\left(\mathcal{R}_{ij}-\frac{\gamma_{ij}}{3}\mathcal{R}\right)-\left(D_iD_j\alpha -\frac{\gamma_{ij}}{3}D_kD^k\alpha\right)\right]+\alpha(K\tilde A_{ij}-2\tilde A_{ik}\tilde A^k_j)\nonumber\\
    &&-\frac{2}{3}(\bar D_k\beta^k)\tilde A_{ij}-\frac{8\pi\alpha}{a^2\psi^4}\left(S_{ij}-\frac{\gamma_{ij}}{3}S_k^k\right),
\end{eqnarray}
\begin{equation}\label{eq31}
    (\partial_t -\mathcal{L}_\beta)\psi = -\frac{\partial_t a}{2a}\psi +\frac{\psi}{6}(-\alpha K+\bar D_k\beta^k),
\end{equation}
\begin{equation}\label{eq_traz_ce}
    (\partial_t -\mathcal{L}_\beta)K = \alpha \left(\tilde A_{ij}\tilde A^{ij}+\frac{1}{3}K^2\right)-D_kD^k\alpha +4\pi\alpha(E+S_k^k),
\end{equation}
with $\mathcal{L}_\beta$ representing the Lie derivative along $\beta^i$.
\end{itemize}

{The evolution} equation {for the spatial metric} \eqref{extcurv} yields
\begin{equation}\label{eq:ren_curv_ext}
    (\partial_t- \mathcal{L}_\beta)\tilde \gamma_{ij} = -2\alpha\tilde A_{ij}-\frac{2}{3}\tilde \gamma_{ij}\bar D_k\beta^k.
\end{equation}

The {Euler-like} system of Eqs. \eqref{fluid1} and \eqref{fluid2} are then rewritten as
\begin{subequations}\label{kgeq}
\begin{equation}\label{eq33a}
    (\partial_t+ v^i\partial_i )\rho +\frac{\rho+p}{(a\psi^2)^3\alpha u^0}\lbrace\partial_t [(a\psi^2)^3\alpha u^0]+\partial_i[(a\psi^2)^3\alpha u^0 v^i]\rbrace = 0,
\end{equation}
\begin{equation}
    \frac{1}{(a\psi^2)^3}\partial_t[(a\psi^2)^3(\rho+p)\alpha u^0 u_i]+D_j[(\rho+p)\alpha u^0 v^ju_i]+\partial_i p+(\rho + p)((\alpha u^0)^2\partial_i \alpha-\alpha u^0 u_jD_i\beta^j) = 0.
\end{equation}
\end{subequations}

The system from \eqref{eq28} to \eqref{kgeq} is then the differential equations that govern the nonlinear evolution for a CSF in a cosmological context {and for the gradient expansion at the order studied in this work}. Observe then that, thanks to the perfect fluid representation of the CSF, all these equations are the same that are valid for a perfect fluid. The particular way to move from the solutions that we will obtain for the perfect fluid variables and the field variables will be presented later in this work.

\subsection{Understanding the cosmological conformal decomposition}\label{subhom}

{Given that this formalism is not the standard for describing cosmological inhomogeneities, in this section we provide a more familiar description of the above system. In particular, we make contact with the standard approach by showing how the well known equations for a cosmological CSF are reproduced at the background level (we leave the treatment of inhomogeneities for the next section, in the context of a gradient expansion).}

{We start by considering a flat Friedmann-Lemaitre metric
\begin{equation}
    ds^2 = -dt^2+a^2(t)\eta_{ij}dx^idx^j.
\end{equation}
This metric fits the 3+1 representation \eqref{metric3+1} with $\alpha = 1$, $\beta^i = 0$, and $\gamma_{ij} = a^2(t)\eta_{ij}$. Furthermore, the cosmological conformal decomposition  takes in this case $\psi = 1$ and $\tilde\gamma_{ij} = \eta_{ij}$. Observe that this also implies that $\mathcal{R}_{ij} = 0$ given that $\mathcal{R}_{ij}^{\psi}$ has only terms of the form $D_i\psi$, and $\mathcal{\tilde R}_{ij}$ can be taken as dependent solely on factors of the form $\hat D_i\eta_{jk}$, which by definition are zero. Additionally, the different quantities in Eq. \eqref{eqs19} reduce to $E = \rho$, $J_i = 0$, and $S_{ij} = pa^2(t)\eta_{ij}$, where we have considered that $u_i = 0$ (the comoving velocity $u_i$ vanishes at the background level) and $\alpha u^0 = 1$, which can be easily seen from Eq. \eqref{equij} once we impose that $\varphi = \varphi(t)$.}

{We now focus on the system of Eqs. \eqref{eq28}-\eqref{kgeq}, which describe the evolution of the CSF in a cosmological context. First, observe that Eq. \eqref{eq:ren_curv_ext} implies that $A_{ij} = 0$. Similarly, Eq. \eqref{eq31} implies that $K = -3\partial_t a/a\equiv -3H$, where $H$ is the well known Hubble parameter. On the other hand, Eq. \eqref{eq28}, which follows from the Hamiltonian constraint, reduces to the Friedmann equation
\begin{equation}\label{eq35q}
    H^2 = \frac{8\pi}{3}\rho,
\end{equation}
whereas Eq. \eqref{eq_traz_ce} reduces to the acceleration equation
\begin{equation}\label{eq36q}
    \frac{\partial_{t}^2a}{a} = -\frac{8\pi}{3}\left[\rho+3 p\right],
\end{equation}
where in the above expression $\partial_{t}^2\equiv \partial^2/\partial t^2$. Similarly, from the mass conservation equation \eqref{eq33a}, it follows that
\begin{equation}\label{eq37q}
    \partial_t \rho +3H(\rho+p) = 0.
\end{equation}
The reader can verify that all the other equations provide no extra information. Thus, Eqs. \eqref{eq35q}-\eqref{eq37q} constitute the system to describe a universe dominated by a CSF at the background level. In fact, it is well known that only two of these equations are independent of each other.  Of course, this system is well known to describe a universe dominated by a perfect fluid, and the reason we arrived at this system is because we are working with the perfect fluid representation of the CSF. In order to close the system of equations, we might be tempted to define an equation of state for the CSF, as is typically done when working with perfect fluids. However, as we mentioned earlier, this fluid representation of the scalar field is only an auxiliary representation, so the definition of an equation of state should be taken with caution. We remark that this problem does not arise in the field representation, where the current conservation equation complements the system (we shall elaborate more about the field variables in the following section). For the meantime we have shown that our formalism is capable of reproducing the usual equations that describe a CSF at the background level. }



\section{Long-wavelength solutions for the complex scalar field}\label{section4}
\subsection{Gradient expansion: Basic assumptions}

The idea behind the gradient expansion formalism \citep{salopek1990nonlinear} is to consider only those configurations with a scale larger than the cosmological horizon size. To this end, we attach to the spatial derivatives a fictitious parameter $\epsilon$, which is typically associated to the scale $L$ of the inhomogeneities as $\epsilon\equiv  H^{-1}/L$. This yields a hierarchy of orders of $\epsilon$ associated to each variable in the system (as shown in Eq.~\eqref{Eq:epsilon_orders} below). Then the system of differential equations is solved order by order in powers of $\epsilon$. 
The formalism, thus, restricts the validity of solutions to scales $L$  much larger than the Hubble horizon, i.e. $L\gg H^{-1}$, so that  $\epsilon\ll 1$ is guaranteed at all times. In this case, the lowest-order terms in $\epsilon$  are sufficient to describe superhorizon inhomogeneities which, on the other hand, are not restricted in amplitude.

Based on the above description, the gradient expansion requires additional considerations. First, it is assumed that $\psi$ acquires the value of one in some asymptotic region of the universe (close to spatial infinity). This makes {$a(t)$ the asymptotic scale factor of the universe and justifies our notation}. A second requirement is  that when $\epsilon\rightarrow 0$ (a value particularly reached at spatial infinity) the universe becomes locally homogeneous and isotropic, i.e., a Friedman universe, which in our case is assumed to be flat. This imposes the asymptotic values $\alpha = O(\epsilon^0)$, $\beta^i = O(\epsilon)$, and $\partial_t{\tilde\gamma}_{ij} = O(\epsilon)$ in the limit $\epsilon\rightarrow 0$,  {(hereafter, $f= O(\epsilon^n)$ means that $f$ is at least of the order $\epsilon^n$ and admits higher-order contributions)}. We guarantee such limits through the more restrictive conditions $\beta^i = O(\epsilon^3)$ and $\partial_t\tilde\gamma_{ij} = O(\epsilon^2)$. The condition $\alpha = O(\epsilon^0)$ implies that $\alpha$ must be a scale-independent quantity at the lowest order in the gradient expansion (consistent with the cosmological solution presented earlier). We can absorb at this order the time-dependence by rescaling the time coordinate without loss of generality and set $\alpha = 1$. On the other hand, imposing $\beta^i = O(\epsilon^3)$ is a matter of coordinate choice, which can be fulfilled in general. Finally, demanding $\partial_t\tilde\gamma_{ij} = O(\epsilon^2)$ is a choice inherited from inflationary solutions. Indeed, as pointed out in Ref. \citep{tanaka2007gradient}, taking $\partial_t\tilde \gamma_{ij} = O(\epsilon)$ introduces a decaying mode at this order, which affects all quantities at $O(\epsilon^2)$. That same reference shows that the condition $\partial_t\tilde \gamma_{ij} = O(\epsilon^2)$ is satisfied for fluctuations arising from vacuum perturbations. Therefore, {given that one of the main motivations of this work is to study perturbations valid for a reheating period}, we expect that such a requirement should be sufficiently general for our {case}.

Since our energy-momentum tensor of the CSF has been expressed in the form of a perfect fluid, the hierarchy in powers of $\epsilon$ of the quantities defined in previous sections is defined from that established for a perfect fluid. That is, by assuming $\beta^i = O(\epsilon^3)$ and $\partial_t\tilde\gamma_{ij} = O(\epsilon^2)$, we have \citep{harada2015cosmological}:
\begin{equation}
\label{Eq:epsilon_orders}
    \bar\Psi = O(\epsilon^0), \ \ \ \ \ \ \ \ \tilde A_{ij},\ f_{ij},\ \kappa,\ \Phi,\ \Psi, \ \partial_t \Psi,\ \delta,\ \delta p=  O(\epsilon^2), \ \ \ \ \ v_i+\beta_i = O(\epsilon^3), \ \ \ \ \ \ \alpha u^0 = 1+O(\epsilon^6).
\end{equation}
where in the above expression $\partial_t\bar\Psi = 0$ and we have defined:
\begin{equation}
     f_{ij} \equiv \tilde\gamma_{ij}-\eta_{ij},\ \ \ \ \ \ \ \ \kappa \equiv \frac{K-\bar K}{\bar K}, \ \ \ \ \ \ \ \ \Phi \equiv \alpha-1, \ \ \ \ \ \ \ \ \delta p \equiv p-\bar p, \ \ \ \ \ \ \ \ \delta \equiv \frac{\rho-\bar\rho}{\bar\rho}, \ \ \ \ \ \ \ \ \psi = \bar\Psi(1+\Psi).
\end{equation}
 
In all these expressions, quantities with a bar are used to refer to $O(\epsilon^0)$, i.e. background, quantities. The above estimations can be translated to the field variables, in which case we obtain 
\begin{equation}
\delta\varphi\equiv \varphi-\bar \varphi = O(\epsilon^2), \qquad \partial_t \delta\varphi = O(\epsilon^2).
\end{equation}
\subsection{Leading-order equations}\label{sec_IVB}

\textit{1. Order $O(\epsilon^0)$:} \\

The first step to construct the leading-order solutions valid for long-wavelength perturbations for a CSF is to reproduce the $O(\epsilon^0)$ equations that govern its evolution. {The reader may find some of  the following expressions redundant with those of Sec. \ref{subhom}; however, we present the results here to be consistent with the gradient expansion formalism, and in order to show the additional equations required to describe the field variables of the CSF.} 

First at all, from Eq. \eqref{eq31} we have 
\begin{equation}
    \bar K = -3H+O(\epsilon^2).
\end{equation}
If now we use the Hamiltonian constraint \eqref{eq28} and Eq. \eqref{Ef}, we obtain
\begin{equation}\label{H_b}
    H^2 = \frac{8\pi}{3}\bar\rho +O(\epsilon^2).
\end{equation}
$\bar\rho$ being written in terms of the field representation as
\begin{equation}\label{rho_b}
\bar\rho = \frac{|\partial_t\bar \varphi|^2}{2}+V(|\bar \varphi|^2){+O(\epsilon^2)},
\end{equation}
and $V(|\bar \varphi|^2)$ is the part of the potential $V(|\varphi|^2)$ that is $O(\epsilon^0)$. From Eqs. \eqref{eq_traz_ce}, \eqref{Ef} and \eqref{Sf} follows
\begin{equation}
     \frac{\partial_{t}^2a}{a} = -\frac{8\pi}{3}\left[\bar\rho+3\bar p\right]+O(\epsilon^2),
\end{equation}
where in the above expression 
$\bar p$ is written in terms of the field variables as 
\begin{equation}\label{p_b}
\bar p = \frac{|\partial_t \bar \varphi|^2}{2}-V(|\bar \varphi|^2){+O(\epsilon^2)}.
\end{equation}
Finally, the fluid equation \eqref{eq33a} is reduced to \begin{equation}\label{back1}
    \partial_t \bar\rho +3H(\bar\rho+\bar p) =O(\epsilon^2).
\end{equation}
If we express the above expression in terms of the field variables, it reduces to
\begin{equation}\label{44d}
    \partial_{t}^2\bar \varphi +3H\partial_t\bar \varphi+2{V'(|\bar \varphi|^2)}\bar \varphi=O(\epsilon^2),
\end{equation}
which is the well known evolution equation for a CSF in a FL background. In the above expression, $V'(|\bar\varphi|^2)$ is the part of $V'(|\varphi|^2)$ that is $O(\epsilon^0)$ \eqref[see Eq.][]{vp}.
To complement the equations at the background level, we need to specify some information on the phase of the scalar field. This information can be obtained from the 4-current conservation equation \eqref{4current}, which at $O(\epsilon^0)$ reads (see appendix \ref{Ap:current}):
\begin{equation}\label{4currentb}
    \partial_t[a^3(\bar \varphi^*\partial_t\bar \varphi-\bar \varphi\partial_t\bar \varphi^*)]=O(\epsilon^2).
\end{equation}
Notice that this last equation is not presented for a RSF (or it is trivially satisfied) and represents a key difference between our scenario and that of a RSF. \\

\textit{2. Order $O(\epsilon^2)$:} \\

Let us continue by presenting the equations that are valid up to $O(\epsilon^2)$. We  adopt the CMC slicing, in which $K=\bar K$ and, thus, $\kappa= 0$.

In the CMC slicing, the system \eqref{eq28}-\eqref{kgeq} is expressed as follows: The Hamiltonian constraint \eqref{eq28} is reduced to
\begin{equation}\label{bardelta}
    \bar \Delta \bar \Psi =-2\pi\bar \Psi^5a^2\bar\rho\delta+O(\epsilon^5).
\end{equation}
{Additionally, the momentum constraint \eqref{momentum} implies that $J_i = O(\epsilon^3)$. While the background Hamiltonian constraint reduces to the Friedmann equation, at the second order it represents a constraint equation for $\bar\Psi$, which encodes deviations from a Friedmann universe.} 

The evolution equations {for the metric variables} are
\begin{equation}\label{eq57}
6\partial_t\Psi-3H\Phi= O(\epsilon^4).
\end{equation}
\begin{equation}\label{eq85}
    \Phi(\bar\rho+\bar p)+\delta p= -\frac{\bar\rho\delta}{3}+O(\epsilon^4).
\end{equation}
\begin{equation}\label{eq62}
        \partial_t f_{ij} = -2\tilde A_{ij} +O(\epsilon^4),
\end{equation}
\begin{equation}\label{eq63}
        \partial_t \tilde A_{ij}+3H\tilde A_{ij} = \frac{1}{a^2\bar \Psi^4}\left[-\frac{2}{\bar \Psi}\left(\bar D_i \bar D_j\bar \Psi -\frac{1}{3}\eta_{ij}\bar\triangle \bar \Psi\right)+\frac{6}{\bar \Psi^2}\left(\bar D_i\bar \Psi\bar D_j\bar \Psi-\frac{1}{3}\eta_{ij}\bar D^k\bar \Psi \bar D_k\bar \Psi\right)\right]+O(\epsilon^4). 
\end{equation}
{Note that Eq.~\eqref{eq85} shows no time derivatives. It is the result of taking the evolution of the extrinsic curvature trace, Eq.~\eqref{eq_traz_ce} in the CMC gauge, where derivatives of the extrinsic curvature trace are zero.} 

Finally, the mass and momentum conservation equations, in terms of the fluid variables, are: 
 
\begin{subequations}\label{density_ev}
\begin{equation}
    \bar\rho\partial_t\delta + (\bar\rho+\bar p)\left(6\partial_t \Psi+D_iv^i\right)+3H(\delta p-\bar p\delta)= O(\epsilon^4),
\end{equation}
and
\begin{equation}
    \frac{1}{a^3}\partial_t \left[a^3(\bar\rho+\bar p)u_i\right]+\partial_i [\delta p+(\bar\rho+\bar p)\Phi] = O(\epsilon^5),
\end{equation}
respectively.
\end{subequations}

Observe that by substituting Eqs. \eqref{back1} and \eqref{eq57} in the above we obtain
\begin{equation}\label{density_2}
    \partial_t(a^2\bar\rho\delta) = O(\epsilon^4),\ \ \ \ \ \ \ \ \ \ \  \frac{1}{a^3}\partial_t[a^3(\bar\rho+\bar p)u_i] = \frac{1}{3}\bar\rho \partial_i\delta +O(\epsilon^5).
\end{equation}

From this equation, we derive, in the next section, the expression for the peculiar velocity. Note that, up to this point, the above set of differential equations is valid for any energy-momentum tensor which can be written in a perfect fluid form. However, {as we mentioned earlier, the hydrodynamic representation of the scalar field is incomplete, so }we must {complement} these equations {with those that apply} to the field variables.

From Eq. \eqref{preassure} we have
\begin{equation}\nonumber
    \delta p = \frac{1}{2}(\partial_t \bar \varphi^*\partial_t\delta\varphi+\partial_t \bar \varphi\partial_t\delta\varphi^*)-\Phi\partial_t\bar \varphi^*\partial_t\bar \varphi-V'(|\bar \varphi|^2)\left(\bar \varphi\delta\varphi^*+\bar \varphi^*\delta\varphi\right)+O(\epsilon^4).
\end{equation}
Using the constraint \eqref{eq85} and substituting Eq.  \eqref{44d} for the background field, we finally obtain
\begin{equation}\label{revarphi}
    \frac{1}{a^3}\partial_t\left[a^3(\partial_t\bar \varphi^*\delta\varphi+\partial_t\bar \varphi\delta\varphi^*)\right] = -\frac{2}{3}\bar\rho\delta+O(\epsilon^4),
\end{equation}
which is a constraint for the field fluctuation. Equivalently, from Eq. \eqref{preassure}, we also have $p = \rho-2V(|\varphi|^2)$ and then
\begin{equation}
    \delta p = \delta \rho -2V'(|\bar \varphi|^2)\left(\bar \varphi\delta\varphi^*+\bar \varphi^*\delta\varphi\right)+O(\epsilon^4).
\end{equation}
Using the above expression and \eqref{eq85} we can express $\Phi$ in terms of the field variables as
\begin{equation}\label{eqeta}
    2V'(|\bar \varphi|^2)[\bar \varphi\delta\varphi^*+\bar \varphi^*\delta\varphi] = \frac{4}{3}\bar\rho\delta+\Phi(\bar\rho+\bar p)+O(\epsilon^4).
\end{equation}

Finally, we can use the 4-current conservation equation \eqref{4current} to obtain a particular constraint for the field variables (see appendix \ref{Ap:current}):
\begin{equation}\label{current_so}
       \partial_t\left\lbrace a^3[\partial_t(\bar \varphi^*\delta\varphi-\bar \varphi\delta\varphi^*)-2(\partial_t\bar \varphi^*\delta\varphi-\partial_t\bar \varphi\delta\varphi^*)]\right\rbrace-a^3\partial_t\left(\frac{\Phi}{a^3}\right)[a^3(\bar \varphi^*\partial_t\bar \varphi-\bar \varphi\partial_t\bar \varphi^*)]=O(\epsilon^4),
\end{equation}
which, again, it is not present in the case of a real SF.

\subsection{Solutions}\label{sec_IVC}

It is well known that for a homogeneous and isotropic CSF there are two well-defined regimes in which the system can be simplified. As pointed out in Refs. \citep{li2014cosmological,charge4,Carvente:2020aae}, for the case in which the oscillations of the CSF are slower than the Hubble expansion ($\omega \ll H$), the CSF is equivalent to a stifflike fluid ($\bar\rho=\bar p$), a behavior expected in the early evolution of the CSF. On the other hand, when the expansion rate of the universe is much larger than the period of oscillations of the CSF ($\omega \gg H$), the CSF is dependent on its particular potential. This last behavior is of particular interest since it is precisely in this regime that CSFs work for different scenarios, (e.g. a reheating phase in the early Universe or a dark matter candidate at later stages). Since the latter regime is of wider interest, we will focus on finding solutions in this fast-oscillating regime, and also comment on the solutions found in the \textit{slow-oscillating} regime. \\

\subsubsection{The fast-oscillating regime ($\omega \gg H$)}

\textit{(a) Order $O(\epsilon^0).-$} Equation \eqref{4currentb} can be solved immediately. For this, it is convenient to rewrite the background CSF $\bar \varphi$ in the polar form:
\begin{equation}\label{phi_b}
    \bar \varphi = |\bar \varphi|e^{i\theta},
\end{equation}
where $|\bar \varphi|$ and $\theta$ are time-dependent functions. Substituting the above expression in \eqref{4currentb}, we obtain 
\begin{equation}\label{current_fo}
    \partial_t\left[2ia^3|\bar \varphi|^2 \omega\right] =\lu{O(\epsilon^2)}.
\end{equation}
Here $\omega\equiv d\theta/dt$ is the angular oscillation frequency of the CSF. Then
\begin{equation}\label{omega1}
    \omega = \frac{Q}{a^3|\bar \varphi|^2}\lu{+O(\epsilon^2)},
\end{equation}
where $Q$ is the charge of the CSF and it is related to the conservation of total number of particles \citep{charge1,li2014cosmological,charge2,charge3,charge4}. {Recall that for a real field the previous equation is trivially satisfied, which means that the RSF particle is also its antiparticle.}

At the background level, we find it more convenient to solve Eq. \eqref{44d} instead of the fluid equation \eqref{back1} {due to the lack of an equation of state, as already argued}. If we replace Eq. \eqref{phi_b} in Eq. \eqref{44d}, we obtain
\begin{equation}\label{kg_b}
     \partial_t^2|\bar\varphi|-|\bar\varphi|\omega^2+3H\partial_t|\bar \varphi|+2V'(|\bar\varphi|^2)|\bar\varphi| =\lu{O(\epsilon^2)}.
\end{equation}
Following the same procedure as in Refs. \citep{li2014cosmological,charge4,Carvente:2020aae}, the fast-oscillating regime applies by demanding  the conditions 
\begin{equation} \label{fast_osc}
\omega\gg H \ \ \ \ \text{and} \ \ \ \ \frac{\partial_t|\bar \varphi|}{|\bar \varphi|}\ll \omega.
\end{equation}
In this case, Eq. \eqref{kg_b} reduces to
\begin{equation}\label{eq_om}
    |\bar \varphi|\omega^2 - 2V'(|\bar \varphi|^2)|\bar \varphi|=\lu{O(\epsilon^2)}.
\end{equation}
Here $\omega = \pm \sqrt{2V'(|\bar \varphi|^2)}\lu{+O(\epsilon^2)}$ account for the two solutions of the oscillation frequency of the CSF. Observe that, from this last expression and Eq. \eqref{omega1}, we have 
\begin{equation}\label{absvarphi}
\pm\sqrt{2V'(|\bar \varphi|^2)}|\bar \varphi|^2 =\frac{Q}{a^3}\lu{+O(\epsilon^2)}, 
\end{equation}
that is, the solution of $|\bar \varphi|$ is given in  terms of the scale factor once $V(|\bar \varphi|^2)$ (and, thus, $V'(|\bar \varphi|^2)$) is specified. With this at hand, we can immediately find the value of $V(|\bar \varphi|^2)$ and $V'(|\bar \varphi|^2)$ in terms of the scale factor. 

From the above equation and \eqref{phi_b}, we have:
\begin{equation}\label{varphi_b}
    \bar \varphi = \frac{C}{[ 2V'(|\bar \varphi|^2)]^{1/4} a^{3/2}}\exp\left[i\int \sqrt{2V'(|\bar \varphi|^2)} dt\right]\lu{+O(\epsilon^2)},
\end{equation}
where $C$ is a complex constant that fulfills the condition $Q =  |C|^2${, and then $C = \sqrt{Q}e^{i\theta_0}$, with $\theta_0$ a global phase}. Also, from Eq. \eqref{phi_b} it follows immediately that 
\begin{equation}\label{pt_varphib}
    \partial_t \bar \varphi =\left(i\sqrt{2V'(|\bar \varphi|^2)}+\frac{\partial_t|\bar\varphi|}{|\bar\varphi|}\right)\bar \varphi. 
\end{equation}

We use the above expressions to calculate the perfect fluid variables. In that case, we obtain from Eqs. \eqref{rho_b} and \eqref{p_b}:
\begin{equation}\label{density_pressure}
    \bar\rho = V'(|\bar \varphi|^2)|\bar \varphi|^2+V(|\bar \varphi|^2)\lu{+O(\epsilon^2)}, \ \ \ \ \ \text{and} \ \ \ \ \ \bar p= V'(|\bar \varphi|^2)|\bar \varphi|^2-V(|\bar \varphi|^2)\lu{+O(\epsilon^2)}.
\end{equation}
This implies that the Hubble parameter $H$ evolves from Eq. \eqref{H_b} as
\begin{equation}\label{H_b2}
    H^2 = \frac{8\pi}{3}\left[V'(|\bar \varphi|^2)|\bar \varphi|^2+V(|\bar \varphi|^2)\right]\lu{+O(\epsilon^2)}.
\end{equation}
{We see that, in the regime of fast oscillations, the dynamics produced by the CSF (the evolution of its energy density or the expansion of the universe) is intimately related to its potential. As we mentioned earlier, this is a well known result.}\\

\textit{Order $O(\epsilon^2)$.-} Given that $\bar\Psi = O(\epsilon^0)$ and $\partial_t \bar\Psi =0$, we have that
\begin{equation}\label{barpsi}
    \bar \Psi = L^{(0)}(x^k),
\end{equation}
where $L^{(0)}(x^k)$ is an arbitrary function of the spatial coordinate $x^k$ and we have used the superscript ${(n)}$ to denote quantities that scale as $\epsilon^n$.
Recall that configurations are subject to a size larger than the horizon.  Equation \eqref{eq63} can be easily solved to obtain
\begin{equation}\label{solAij}
    \tilde A_{ij} = p^{(2)}_{ij}\frac{1}{a^3}\int_0^a\frac{d\tilde a}{H(\tilde a)}+O(\epsilon^4),
\end{equation}
where 
\begin{equation}\nonumber
    p^{(2)}_{ij}(x^k) \equiv \frac{1}{\bar \Psi^4}\left[-\frac{2}{\bar \Psi}\left(\bar D_i \bar D_j\bar \Psi -\frac{1}{3}\eta_{ij}\bar\triangle \bar \Psi\right)+\frac{6}{\bar \Psi^2}\left(\bar D_i\bar \Psi\bar D_j\bar \Psi-\frac{1}{3}\eta_{ij}\bar D^k\bar \Psi \bar D_k\bar \Psi\right)\right],
\end{equation}
and with $H$ given by Eq. \eqref{H_b}. The integration constant in Eq.~\eqref{solAij} was omitted, since it yields a decaying mode, irrelevant for our purposes. Equation \eqref{eq62} yields immediately
\begin{equation}\label{solhij}
    f_{ij} = -2\int_0^a\frac{ d\tilde a A_{ij}(\tilde a)}{\tilde a H(\tilde a)}+O(\epsilon^4),
\end{equation}
where the constant of integration is dropped in order that $f_{ij}\rightarrow 0$ when $t\rightarrow 0$. 

Next, the conservation equation \eqref{density_2} is easily integrated. The density is integrated as
\begin{equation}\label{delta}
    \delta = \frac{\bar\rho_0a_0^2}{\bar\rho a^2}R^{(2)}(x^k)+O(\epsilon^4),
\end{equation}
with $R^{(2)}(x^k)$ an arbitrary function of the spatial coordinates $x^k$, {restricted only to represent a fluctuation of size larger than the cosmological horizon,} and with $\bar\rho_0a_0^2$ a constant. Using the above expression, {the right-hand side} equation in \eqref{density_2} yields the peculiar velocity:
\begin{equation}\label{solui}
    u_i =\frac{\bar\rho_0a_0^2}{3a^3(\bar\rho+\bar p)}\partial_i R^{(2)}(x^k)\int_0^a \frac{d\tilde a}{H(\tilde a)}\lu{+O(\epsilon^3)}.
\end{equation}

The functions $L^{(0)}(x^k)$ and $R^{(2)}(x^k)$ are related by Eq. \eqref{bardelta} as
\begin{equation}\label{eqRL}
    R^{(2)}(x^k) = -\frac{\bar \Delta L^{(0)}(x^k)}{2\pi\bar\rho_0a_0^2 (L^{(0)}(x^k))^5}\lu{+O(\epsilon^5)}.
\end{equation}

Let us now solve Eq. \eqref{revarphi} with the aid of Eq. \eqref{delta}. In this case, we obtain
\begin{equation}\label{eq:1}
    \partial_t\bar \varphi^*\delta\varphi+\partial_t \bar \varphi\delta\varphi^* = -\frac{2\bar\rho_0a_0^2}{3a^3}R^{(2)}(x^k)\int_0^a\frac{d\tilde a}{H(\tilde a)}+O(\epsilon^4).
\end{equation}
Observe that in the fast-oscillating regime we can approximate  $\partial_t \bar \varphi\simeq i\sqrt{2V'(|\bar\varphi|^2)} \bar \varphi$, from Eq. \eqref{pt_varphib}. Then 
\begin{equation}\label{difvarphi}
    \bar \varphi^*\delta\varphi- \bar \varphi\delta\varphi^* \simeq -\frac{i}{\sqrt{2V'(|\bar \varphi|^2)}}\frac{2\bar\rho_0a_0^2}{3a^3}R^{(2)}(x^k)\int_0^a\frac{d\tilde a}{H(\tilde a)}+O(\epsilon^4).
\end{equation}
Equivalently, from Eq. \eqref{eqeta} we have
\begin{equation}\label{diffbarphit}
    \partial_t\bar \varphi^*\delta\varphi-\partial_t\bar \varphi\delta\varphi^* \simeq -\frac{i}{\sqrt{2V'(|\bar \varphi|^2)}}\left[\frac{4}{3}\frac{\bar\rho_0a_0^2}{a^2}R^{(2)}(x^k)+\Phi(\bar\rho+\bar p)\right]+O(\epsilon^4).
\end{equation}
Using these two expressions and Eq. \eqref{current_fo} in Eq. \eqref{current_so}, we obtain immediately
\begin{eqnarray}\label{eqapb}
    &&\partial_t \left\lbrace a^3\left[\partial_t\left(\frac{1}{\sqrt{2V'(|\bar \varphi|^2)}}\frac{2\bar\rho_0a_0^2}{3a^3}R^{(2)}(x^k)\int_0^a\frac{d\tilde a}{H(\tilde a)}\right)-\frac{2}{\sqrt{2V'(|\bar \varphi|^2)}}\left(\frac{4}{3}\frac{\bar\rho_0a_0^2}{a^2}R^{(2)}(x^k)+\Phi(\bar\rho+\bar p)\right)\right]\right\rbrace\nonumber \\
    &&+2Qa^3\partial_t\left(\frac{\Phi}{a^3}\right) = O(\epsilon^4).
\end{eqnarray}
By noticing from Eqs. \eqref{density_pressure} and \eqref{absvarphi} that $\bar\rho+\bar p \simeq 2V'(|\bar \varphi|^2)|\bar\varphi|^2= \sqrt{2V'(|\bar\varphi|^2)}Q/a^3$, and using $\partial_t = aH\partial_a$, the above expression is rewritten as
\begin{eqnarray}
    &&aH\partial_a\left\lbrace a^3\left[aH\partial_a\left(\frac{1}{\sqrt{2V'(|\bar \varphi|^2)}}\frac{2\bar\rho_0a_0^2}{3a^3}R^{(2)}(x^k)\int_0^a\frac{d\tilde a}{H(\tilde a)}\right)-\frac{2}{\sqrt{2V'(|\bar \varphi|^2)}}\frac{4}{3}\frac{\bar\rho_0a_0^2}{a^2}R^{(2)}(x^k)\right]\right\rbrace\nonumber \\
    &&+2Q\partial_t\Phi-2Q\partial_t\Phi-6QH\Phi = \lu{O(\epsilon^4)}. 
\end{eqnarray}
Then
\begin{equation}\label{eq:Phi}
    \Phi \simeq \frac{a}{6Q}\partial_a \left\lbrace a^3\left[aH\partial_a\left(\frac{1}{\sqrt{2V'(|\bar \varphi|^2)}}\frac{2\bar\rho_0a_0^2}{3a^3}R^{(2)}(x^k)\int_0^a\frac{d\tilde a}{H(\tilde a)}\right)-\frac{2}{\sqrt{2V'(|\bar \varphi|^2)}}\left(\frac{4}{3}\frac{\bar\rho_0a_0^2}{a^2}R^{(2)}(x^k)\right)\right]\right\rbrace+O(\epsilon^4).
\end{equation}
We can use this last result to solve Eq. \eqref{eq57}:
\begin{equation}\label{eq:Psi}
    \Psi = \frac{1}{2}\int_0^a\Phi\frac{d\tilde a}{\tilde a} =  \frac{a^3}{12Q}\left[aH\partial_a\left(\frac{1}{\sqrt{2V'(|\bar \varphi|^2)}}\frac{2\bar\rho_0a_0^2}{3a^3}R^{(2)}(x^k)\int_0^a\frac{d\tilde a}{H(\tilde a)}\right)-\frac{2}{\sqrt{2V'(|\bar \varphi|^2)}}\left(\frac{4}{3}\frac{\bar\rho_0a_0^2}{a^2}R^{(2)}(x^k)\right)\right]+O(\epsilon^4).
\end{equation}
Finally, by adding Eqs. \eqref{eq:1} and \eqref{diffbarphit}, we can obtain the solution for $\delta\varphi$:
\begin{equation}\label{varphip}
    \delta\varphi =- \frac{1}{\partial_t\bar\varphi^*}\left[\frac{2\bar\rho_0a_0 ^2}{3a^3}R^{(2)}(x^k)\left(\int_0^a\frac{d\tilde a}{H(\tilde a)}+\frac{i2a}{\sqrt{2V'(|\bar\varphi|^2)}}\right)+\frac{i\Phi(\bar\rho+\bar p)}{\sqrt{2V'(|\bar \varphi|^2)}}\right]\lu{+O(\epsilon^4),}
\end{equation}
where $\Phi$ is given by Eq. \eqref{eq:Phi}. 

\subsubsection{The slow-oscillating regime ($\omega \ll H$)}

\textit{Order $O(\epsilon^0).-$} In this regime, the solution  obtained in Eq. \eqref{omega1} holds valid given that Eq. \eqref{current_fo} is general and independent of any approximation. On the other hand, following again Refs. \citep{li2014cosmological,charge4,Carvente:2020aae}, in the slow-oscillating regime we consider $\omega \ll H$ and $\partial_t |\bar\varphi|/|\bar\varphi|\gg \omega$, in which case, Eq. \eqref{kg_b} is rewritten as
\begin{equation}
    \partial_t^2|\bar\varphi|+3H\partial_t |\bar\varphi| = \lu{O(\epsilon^2)}.
\end{equation}
The above expression can be immediately integrated to obtain,
\begin{equation}\label{pt_ph}
    \partial_t |\bar\varphi| = \partial_t |\bar\varphi|_0\left(\frac{a_0}{a}\right)^3\lu{+O(\epsilon^2)},
\end{equation}
where $\partial_t |\bar\varphi|_0$ is a constant. 

The energy and pressure in this case are given by
\begin{equation}
    \bar\rho \simeq \bar p \simeq \frac{(\partial_t|\bar\varphi|)^2}{2}\lu{+O(\epsilon^2)} = \frac{(\partial_t|\bar\varphi|_0)^2}{2}\left(\frac{a_0}{a}\right)^6\lu{+O(\epsilon^2)}.
\end{equation}
The above equation yields the equation of state of a stifflike fluid (defined only for background quantities), which is well known to apply for a CSF at the earliest epoch and is independent of the particular potential of the CSF, {since at this stage it is the kinetic energy of the scalar field particles (the first term in the quantities defined in Eq. \eqref{preassure}) that dominates the energy density}. As pointed out in Ref. \citep{li2014cosmological}, this stifflike behavior of the CSF implies that the sound speed associated to it almost reaches the speed of light (the maximum value allowed), which is an analog to the incompressible
fluid in Newtonian gas dynamics, where the sound speed is infinite. Substituting the above expression in Eq. \eqref{H_b}, we obtain the Hubble parameter $H$ evolution as
\begin{equation}\label{h2st}
    H^2 = \frac{4\pi(\partial_t|\bar\varphi|_0)^2}{3}\left(\frac{a_0}{a}\right)^6\lu{+O(\epsilon^2)}.
\end{equation}
Using the above equation in Eq. \eqref{pt_ph}, we obtain an expression for $|\bar\varphi|$, namely,
\begin{equation}
    |\bar\varphi| = |\bar\varphi|_0 +\sqrt{\frac{3}{4\pi}}\ln\left(\frac{a}{a_0}\right)\lu{+O(\epsilon^2)}.
\end{equation}

It is easy to see that in the slow-oscillating regime we will have that $\partial_t\bar\varphi = e^{i\theta}\partial_t|\bar\varphi|$, so to complete this subsection at the background level we need to calculate the value of the phase of the CSF. We can do this easily with the help of Eq. \eqref{omega1} and the above two expressions as follows:
\begin{equation}
     \theta -\theta_0= \int_0^a \omega \frac{d\tilde a}{\tilde a H(\tilde a)} = -\frac{Q}{a_0^3\partial_t|\bar\varphi|_0}\frac{1}{|\bar\varphi|_0+\sqrt{3/(4\pi)}\ln(a/a_0)}\lu{+O(\epsilon^2)},
\end{equation}
where $\theta_0$ is {the global constant phase defined in the paragraph just after Eq. \eqref{varphi_b}}. Notice that the above result can be expressed as $\theta-\theta_0 \sim -\omega |\bar\varphi|/H$, and, given that $H\sim \partial_t|\bar\varphi|$, we have $\theta-\theta_0\sim -\omega |\bar\varphi|/\partial_t|\bar \varphi|\ll 1$, since we are in the slow-oscillating regime. This last result then allows us to approximate 
\begin{equation}\label{bk_st}
    \bar\varphi \simeq |\bar\varphi|e^{i\theta_0}, \ \ \ \ \ \ \ \partial_t\bar\varphi \simeq \partial_t|\bar\varphi|e^{i\theta_0}.
\end{equation}
This is consistent with the physical picture of the slow-oscillating regime, in which the
Hubble time is much smaller than the oscillation
period, so that the CSF rolls down the potential well, before
completing a cycle of spin.

{It is worth mentioning that in the slow-oscillating regime, a RSF would be expected to feature a subsequent attractor solution of an effective cosmological constant \citep{piran1985inflation} (see also Refs. \citep{charge4,Padilla:2019fju}). For a CSF, however, such behavior demands specific conditions, so the inflationary models that would arise from a CSF are not as generic as in the case of real fields.} \\

\textit{Order $O(\epsilon^2)$.-} {Before presenting the solutions for this regime at second order, the reader should note that some of the solutions found in the fast-oscillating regime are also valid in the present case. This is because the background solutions are the same up to time derivatives of the field. Such factors are not present in the solutions provided by  equations \eqref{barpsi} to \eqref{eq:1}. As for the rest of the solutions, we note that by using \eqref{pt_ph}, \eqref{h2st} and \eqref{bk_st}, equation \eqref{eq:1} can be expressed as}
\begin{equation}
    e^{-i\theta_0}\delta\varphi+e^{i\theta_0}\delta\varphi^* = -\frac{1}{4\sqrt{12\pi}}\left(\frac{a}{a_0}\right)^4R^{(2)}(x^k)+O(\epsilon^4),
\end{equation}
where in the above expression we have used that $\bar\rho_0 = (\partial_t |\bar\varphi|_0)^2/2$. Using the last equation, we can solve $\Phi$ by using Eq. \eqref{eqeta}. In such a case, we obtain
\begin{equation}
    \Phi = -\frac{R^{(2)}(x^k)}{(\partial_t |\bar\varphi|)^2}\left(\frac{a}{a_0}\right)^6\left[\frac{1}{4\sqrt{12\pi}}\left(\frac{a}{a_0}\right)^4 2V'(|\bar\varphi|^2)\left(|\bar\varphi|_0+\sqrt{\frac{3}{4\pi}}\ln\left(\frac{a}{a_0}\right)\right)+\frac{2}{3}\frac{(\partial_t|\bar\varphi|_0 )^2 a_0^2}{a^2}\right]+O(\epsilon^4).
\end{equation}
It is interesting to notice that the value of $\Phi$ will depend on the particular value of the potential under which the CSF is subject. Then, from Eq. \eqref{eq57}, we have
\begin{equation}
    \Psi = \int_0^a \Phi\frac{d\tilde a}{\tilde a} \lu{+O(\epsilon^4)}= - \frac{R^{(2)}(x^k)}{(\partial_t |\bar\varphi|)^2}\int_0^a\frac{\tilde a^5}{a_0^6}\left[\frac{1}{4\sqrt{12\pi}}\left(\frac{\tilde a}{a_0}\right)^4 2V'(|\bar\varphi|^2)\left(|\bar\varphi|_0+\sqrt{\frac{3}{4\pi}}\ln\left(\frac{\tilde a}{a_0}\right)\right)+\frac{2}{3}\frac{(\partial_t|\bar\varphi|_0 )^2 a_0^2}{\tilde a^2}\right]d\tilde a+O(\epsilon^4).
\end{equation}

While the slow-oscillating regime is not usually featured in reheating models, it is plausible to consider a stiff-fluid component at early times given its dependence with the scale factor. In such scenario, the formation of PBHs has been studied in previous papers (e.g. Refs. \citep{Sahni:2001qp,Hidalgo:2011fj}) but with no rigorous criterion for the threshold amplitude for their formation. The present study serves as a first step in the determination of such an amplitude.

{To close this section, let us note from Eq.~\eqref{eqRL} that all inhomogeneities present a spatial dependence related to  $R^{(2)}(x^k)$ and its derivatives. This means that a single spatial distribution governs all quantities in the system. This is consistent with the picture in which our solutions take into account only the growing mode, which is the case of structure formation preceded by an inflationary period which erased all decaying modes. In particular, keeping exclusively this mode is crucial in the study of the formation of PBHs since the incorporation of the decaying mode brings uncertainties to the determination of the threshold amplitude (at horizon crossing), with which an overdensity may collapse gravitationally and form a PBH.}

\section{Characterization of two explicit potentials}\label{section5}

We have written the Einstein equations in Sec. \ref{sec_IVB} in the gradient expansion approximation  and presented solutions for a generic canonical potential in Sec. \ref{sec_IVC} for both the fast- and the slow-oscillating regime.  We now proceed to derive explicit solutions for some particular potentials, namely, a quadratic and a quartic potential. Since the fast-oscillating regime is of most interest in cosmology, we will focus on studying examples only in this regime (considering also that solutions in the regime of slow oscillations are immediately recovered by simply substituting the quantities in the general solutions presented above).

{Our choice of potentials responds to our focus on cases of interest for reheating, where our solution is suitable to model the evolution of perturbations generated during inflation and that lie outside the cosmological horizon.} {The idea is that immediately after the end of inflation, the CSF quickly rolled down to its minimum, where, in the case of large field models, the potential is approximated as $V(|\varphi|^2)\simeq C_n^2|\varphi|^{2n}$, with $C_n$ a suitable constant. With this in mind, we shall analyze the cases $n=1$ and $n=2$. Also, for our purposes we must remember that inflation typically ends when $\epsilon \equiv \frac{M_{pl}}{2}\left(\frac{dV/d\varphi}{V}\right)^2\simeq 1$. Assuming for simplicity that the transition from the domain of the inflationary behavior to the behavior of oscillations around the minimum is instantaneous, we can use the above potential in the end-of-inflation condition, which means that inflation ends when $\varphi\sim M_{pl}$ {and} $C_n\sim H$.
Subsequently, at some point soon after the end of inflation, 
\begin{equation}\label{C_n}
C_n\gg H.
\end{equation}
Then the CSF experiences fast oscillations around the minimum of its potential until interaction with other fields results in its decay to standard model fields. 
}

\subsection{The quadratic potential}

We first consider the following simple harmonic potential:
\begin{equation}\label{potential}
    V(|\varphi|^2) = \frac{\mu^2}{2}|\varphi|^2,
\end{equation}
where $C_1 = \mu^2/2$. This can be used to describe a mass term associated to the CSF particles, and this potential is the minimum necessary that is usually used for the CSF to have a dustlike behavior at late times, so it is precisely this potential that is used to consider the CSF as a candidate for dark matter or to describe a type of reheating process. Observe that in this case 
\begin{equation}
V(|\bar\varphi|^2)= \frac{\mu^2}{2}|\bar\varphi|^2 \ \ \ \ \ \ \text{and} \ \ \ \ \ \ V'(|\varphi|^2)= \mu^2 = V'(|\bar\varphi|^2).
\end{equation} 
{Then, in this example the fast-oscillating regime is fulfilled when the condition $\mu \gg H$ applies, {which is equivalent to the condition in Eq. \eqref{C_n}}.}\\

\textit{1. Order $O(\epsilon^0):$}\\

We can start by finding the solutions that are valid to $O(\epsilon^0)$. Observe that from Eq. \eqref{eq_om} we have $\omega =\pm \mu$. From Eq. \eqref{absvarphi}, this yields
\begin{equation}
    |\bar\varphi|^2 = \pm\frac{Q}{\mu a^3}\lu{+O(\epsilon^2)}.
\end{equation}
Equation \eqref{varphi_b} follows immediately:
\begin{equation}
    \bar \varphi = \frac{C}{\mu^{1/2} a^{3/2}}e^{i\mu t}\lu{+O(\epsilon^2)}.
\end{equation}

{The frequency $\mu$ brings naturally an associated characteristic length scale, which  appears explicitly in the stability analysis of linear perturbations, namely, the instability scale $l_{\rm inst} \approx a / \mu $, which divides fluctuation sizes into two regimes.
For inhomogeneities of size $L \gg l_{\rm inst}$, the background behaves like pressureless dust and perturbations can grow without limit \cite{Jetzer,Carrion:2021yeh}, while in the opposite regime, the scalar field fluctuations dilute and are, therefore, irrelevant for structure formation. Note that, since  $l_{\rm inst} \ll 1/H$, the configurations in the long-wavelength approximation lie well within the regime relevant for structure formation
\footnote{The linear instability scale is found in the analysis of the Mukhanov-Sasaki equation and is often dubbed the Jeans instability of the scalar field \cite{Jetzer,Carrion:2021yeh}. The growth of inhomogeneities above the instability scale is observed in the nonlinear regime both through numerical simulations (with the formation of soliton structures, e.g. \cite{sfdmrh2}) and in the long-wavelength approximation of nonlinear fluctuations in a real scalar field \cite{Sasaki:1998ug}.}.} 
                                      
{From the above equation we have}
\begin{equation}
    \partial_t \bar\varphi =\mu\left(i-\frac{3}{2}\frac{H}{\mu}\right)\bar\varphi\simeq i\mu \bar\varphi. 
\end{equation}
We substitute these expressions in Eqs. \eqref{rho_b} and \eqref{p_b} or in Eq. \eqref{density_pressure}, to obtain
\begin{equation}\label{density_pressure_p2}
    \bar\rho \simeq \frac{\mu Q}{a^3}\lu{+O(\epsilon^2)}, \ \ \ \ \ \text{and} \ \ \ \ \ \bar p\simeq \lu{O(\epsilon^2)}.
\end{equation}
This implies that the Hubble parameter ${H}$ evolves from Eq. \eqref{H_b} or \eqref{H_b2} as
\begin{equation}\label{H_b3}
    {H}^2 = \frac{8\pi}{3}\frac{\mu Q}{a^3}\lu{+O(\epsilon^2)},
\end{equation}
i.e., as a dustlike component, as expected.

{The constants $Q$ and $C$ are determined in each specific example.  In our case, a reheating scenario, we assume that reheating started immediately after the end of inflation. Then  Eq. \eqref{H_b3} sets
\begin{equation}\label{eq_for_q}
    Q = \frac{3 H_{0}^2}{8\pi\mu}a_{0}^3,
\end{equation}
where  the subscript ``$0$" refers to quantities evaluated at the end of inflation. Correspondingly, $C$ is given by
\begin{equation}\label{eq_for_c}
    C = \sqrt{\frac{3 H_{0}^2a_{0}^3}{8\pi\mu}}e^{i\theta_0}.
\end{equation}
}

\textit{2. Order $O(\epsilon^2):$}\\

Now, we proceed by finding the solutions that are valid up to the order $O(\epsilon^2)$. First of all, by substituting our background solutions in Eq. \eqref{solAij}, we obtain
\begin{equation}\label{A_ijge}
    \tilde A_{ij} = p_{ij}^{(2)}\sqrt{\frac{3}{2\pi \mu Q}}\frac{1}{5a^{1/2}}+O(\epsilon^4).
\end{equation}
Replacing the above expression into Eq. \eqref{solhij} results in
\begin{equation}
    f_{ij} = -p_{ij}^{(2)}\frac{3}{10\pi \mu Q}a+O(\epsilon^4).
\end{equation}

We can also substitute our background quantities in Eqs. \eqref{delta} and \eqref{solui}, in which case we have
\begin{equation}\label{delta_dust}
    \delta = \frac{a}{a_0}R^{(2)}(x^k)+O(\epsilon^4), \ \ \ \ \ \ \ \ u_i =\sqrt{\frac{1}{6\pi \mu Q}}\frac{a^{5/2}}{5a_0}\partial_i R^{(2)}(x^k)+O(\epsilon^5),
\end{equation}
where in the above expression we substituted $\bar\rho_0a^2_0 = \mu Q/a_0$. 

{Fig. \ref{fig1} we show the evolution of the above two solutions in terms of the scale factor, normalized at the end of inflation.} Note that, as expected, the density contrast and velocity of the CSF subject to a potential of the form \eqref{potential} evolve as those of a dustlike component \citep{harada2015cosmological}. 

The next step is to find $\Phi$ from Eq. \eqref{eq:Phi}. In that case, we found
\begin{equation}
    \Phi = -\frac{7}{15}\frac{a}{a_0}R^{(2)}(x^k)+O(\epsilon^4).
\end{equation}
If we use this result in Eq. \eqref{eq:Psi}, it follows that
\begin{equation}
    \Psi = -\frac{7}{30}\frac{a}{a_0}R^{(2)}(x^k)+O(\epsilon^4).
\end{equation}

Finally, we compute the solution of the CSF perturbation \eqref{varphip}. In this case, we obtain:
\begin{equation}
    \delta\varphi = \left[\frac{13}{15}\frac{\sqrt{Q}}{\sqrt{\mu}a_0\sqrt{a}}-i\frac{4}{15}\sqrt{\frac{3}{8\pi }} \frac{a}{a_0}\right]R^{(2)}(x^k)e^{i\mu t+i\theta_0}+O(\epsilon^4).
\end{equation}
{If we use Eqs. \eqref{eq_for_q} and \eqref{eq_for_c}, the above expression is reduced to
\begin{equation}
    \delta\varphi = \frac{1}{15}\sqrt{\frac{3}{8\pi}}\left[13 \frac{H_0}{\mu}\sqrt{\frac{a_0}{a}}-i4\frac{a}{a_0}\right]R^{(2)}(x^k)e^{i\mu t +\theta_0}\lu{+O(\epsilon^4)}.
\end{equation}

Fig. \ref{fig1} we  show the evolution of the real part of the field fluctuation as a function of the scale factor, with the numerical value $\mu/H_0 = 10$ (which fulfills the fast-oscillating condition $\mu\gg H$) and $t_0 = 2/(3H_0)$. In the figure, we have multiplied $\delta\varphi$ for this case by a factor of $10$ to ease the comparison with the quartic case presented below. As expected, the CSF experiences fast oscillations in both the background and its fluctuation (the reader may, in fact, verify that the oscillation frequency of $\delta\varphi$ coincides with the one for $\bar\varphi$).} {In addition to the oscillating behavior, the value of the amplitude of $\delta\varphi$ grows with the scale factor. This is, of course, a hallmark of the growing mode of the CSF configuration.}
\subsection{The quartic potential}

We now consider the quartic potential
\begin{equation}\label{potential_4}
    V(|\bar\varphi|^2) = \frac{\lambda}{4}|\varphi|^4,
\end{equation}
with $\lambda>0$. Then
\begin{equation}\label{eqvquartic}
    V(|\bar\varphi|^2)= \frac{\lambda}{4}|\bar\varphi|^4, \ \ \ \ \ V'(|\varphi|^2)= \frac{\lambda}{2}|\varphi|^2 \ \ \ \ \ \ \ \text{and}\ \ \ \ \ \ \  V'(|\bar\varphi|^2) = \frac{\lambda}{2}|\bar\varphi|^2.
\end{equation}
{The fast-oscillating behavior in this case, is guaranteed as long as the condition $\sqrt{\lambda}|\bar\varphi|\gg H$ holds.}

It is well known that this potential undergoes a radiationlike era in its fast-oscillating regime \citep{li2014cosmological,charge4}. It has been used as an intermediate epoch for the SFDM, that is, a potential term like this could dominate before the mass term does, or it could take place in the reheating process. In this sense, our results may be valid for potentials which contain both terms, Eqs.~\eqref{potential} and \eqref{potential_4}, in the limit of large values of $\varphi$.\\

\textit{1. Order $O(\epsilon^0):$}\\

Following the same procedure as with the quadratic case, we start by finding the solutions that are valid at zeroth order. Observe that from Eq. \eqref{eq_om} we have $\omega =\pm \sqrt{\lambda}|\bar\varphi|$, and then, from Eq. \eqref{absvarphi}, we find
\begin{equation}\label{absphi4}
    |\bar\varphi| =\pm \frac{(Q/\sqrt{\lambda})^{1/3}}{a}\lu{+O(\epsilon^2)}.
\end{equation}
We use the above expression and Eq. \eqref{eqvquartic} in Eq. \eqref{density_pressure} to obtain
\begin{equation}\label{density_pressure_p4}
    \bar\rho \simeq \frac{3}{4}\frac{(\lambda Q^4)^{1/3}}{a^4}\lu{+O(\epsilon^2)}, \ \ \ \ \ \text{and} \ \ \ \ \ \bar p\simeq \frac{1}{4}\frac{(\lambda Q^4)^{1/3}}{a^4}\lu{+O(\epsilon^2)}.
\end{equation}
Observe that from the above expression $ \bar p =\bar\rho/3 $, which thus mimics a radiationlike fluid. This implies that the Hubble parameter ${H}$ evolves from Eq. \eqref{H_b} or \eqref{H_b2} as
\begin{equation}\label{H_b4}
    {H}^2 = {2\pi}\frac{(\lambda Q^4)^{1/3}}{a^4}\lu{+O(\epsilon^2)}.
\end{equation}

Using the above expressions, we can finally solve for the homogeneous scalar field. If we replace Eqs. \eqref{eqvquartic} and \eqref{absphi4} in Eq. \eqref{varphi_b}, we get
\begin{equation}
    \bar\varphi = \frac{ C}{(\lambda Q)^{1/6}a}\exp\left[i(\lambda Q)^{1/3}\int_0^a \frac{d\tilde a}{\tilde a^2 H(\tilde a)}\right]\lu{+O(\epsilon^2)}.
\end{equation}
Substituting Eq. \eqref{H_b4} in the above expression, we obtain
\begin{equation}
    \bar\varphi = \frac{ C}{(\lambda Q)^{1/6}a}\exp\left[i\frac{1}{\sqrt{2\pi}}\left(\frac{\sqrt{\lambda}}{Q}\right)^{1/3}a\right]\lu{+O(\epsilon^2)}.
\end{equation}

Considering a reheating stage, the values of $Q$ and $C$ set the initial conditions at the end of inflation:
\begin{equation}
\label{qandcs}
    Q = \left(\frac{a_0^4 H_0^2}{2\pi \lambda^{1/3}}\right)^{3/4}, \ \ \ \ \ \ \ \ \ \ C = \left(\frac{a_0^4 H_0^2}{2\pi \lambda^{1/3}}\right)^{3/8}e^{i\theta_0}.
\end{equation}

\textit{2. Order $O(\epsilon^2):$}\\

Let us now find the solutions that are valid up to the order $O(\epsilon^2)$. By substituting Eq. \eqref{H_b4} in Eq. \eqref{solAij}, we obtain
\begin{equation}\label{A_ijge2}
    \tilde A_{ij} = p_{ij}^{(2)}\frac{1}{\sqrt{2\pi}(\lambda Q)^{1/6}}+O(\epsilon^4).
\end{equation}
Equation \eqref{solhij} yields immediately
\begin{equation}
    f_{ij} = -p_{ij}^{(2)}\frac{a^2}{\pi(\lambda Q^4)^{1/3}}+O(\epsilon^4).
\end{equation}

Substituting our background quantities \eqref{density_pressure_p4} in Eqs. \eqref{delta} and \eqref{solui} gives
\begin{equation}\label{delta_rad}
    \delta = \frac{a^2}{a_0^2}R^{(2)}(x^k)+O(\epsilon^4), \ \ \ \ \ \ \ \ u_i =\frac{1}{4\sqrt{2\pi}(\lambda Q^4)^{1/6}}\frac{a^{4}}{a_0^2}\partial_i R^{(2)}(x^k)+O(\epsilon^5),
\end{equation}
where we used $\bar\rho_0a^2_0 = 3(\lambda Q^4)^{1/3}/(4a_0^2)$. 

\begin{figure}
    \centering
    \includegraphics[width=3.4in]{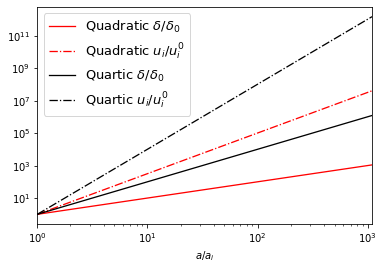}\ \
    \includegraphics[width=3.4in]{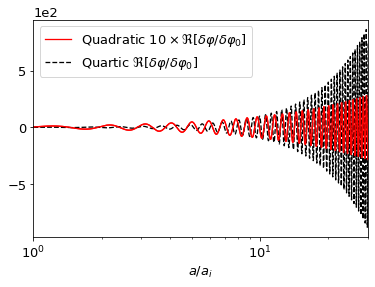}
    \caption{{Left: normalized contrast density and velocity fields for the quadratic and quartic scenarios in terms of the scale factor. Right: evolution of the real parts of $\delta\varphi/\delta\varphi_0$ for the quadratic and quartic scenarios in terms of the scale factor. In the plots $\delta_0$, $u_i^0$, and $\delta\varphi_0$ are the contrast density, velocity, and CSF inhomogeneity, respectively, measured at the end of inflation.}}
    \label{fig1}
\end{figure}

{In Fig. \ref{fig1} we show the evolution of the two  expressions above as a function of the scale factor, normalized at the end of inflation. As mentioned earlier, the density contrast and velocity of the CSF subject to a quartic potential evolve like a pure radiation fluid \citep{harada2015cosmological}, which results in a faster growth with respect to the quadratic case. {This implies that the size of the perturbations (and, in general, their behavior) when they enter the Hubble horizon will be strongly affected by its potential. In the specific case of our two examples, for a common initial power spectrum, we expect a CSF subject to a quartic potential to have a larger value of the amplitude of the overdensity when reentrering the Hubble horizon than for the massive field case.}}

Let us finally derive $\Phi$ from Eq. \eqref{eq:Phi}. In this particular case, we find
\begin{equation}
    \Phi \simeq -\frac{1}{2}\frac{a^2}{a_0^2}R^{(2)}(x^k)\lu{+O(\epsilon^4)}.
\end{equation}
If we use this result in Eq. \eqref{eq:Psi}, then
\begin{equation}
    \Psi = -\frac{1}{8}\frac{a^2}{a_0^2}R^{(2)}(x^k)\lu{+O(\epsilon^4)}.
\end{equation}

Thus, the solution of the CSF perturbation up to the order of $O(\epsilon^2)$ is, from Eq. \eqref{varphip},
{\begin{equation}
    \delta \varphi = \left[\frac{1}{2}\left(\frac{Q^2}{\lambda}\right)^{1/6}\frac{1}{aa_0^2}-\frac{i}{2\sqrt{2\pi}}\frac{a^2}{a_0^2}\right]R^{(2)}(x^k)\exp\left[i\frac{1}{\sqrt{2\pi}}\left(\frac{\sqrt{\lambda}}{Q}\right)^{1/3}a+i\theta_0\right]+O(\epsilon^4).
\end{equation}
Using the parameters in Eq. \eqref{qandcs}, the above expression is 
\begin{equation}
    \delta\varphi = \left[\frac{1}{2}\left(\frac{H_0^2}{2\pi \lambda}\right)^{1/4}\frac{a_0}{a}-i\frac{1}{2\sqrt{2\pi}}\left(\frac{a}{a_0}\right)^2\right]R^{(2)}(x^k)\exp\left[i\frac{1}{\sqrt{2\pi}}\left(\frac{2\pi\lambda}{H_0^2}\right)^{1/4}\frac{a}{a_0}+i\theta_0\right]\lu{+O(\epsilon^4).}
\end{equation}
In Fig. \ref{fig1} we show the evolution of {the real part of} $\delta\varphi$ as a function of the scale factor. The fast-oscillating condition is fulfilled as long as $\left(2\pi H^2/\lambda\right)^{1/4}\ll 1$, {which turns out to be the condition imposed in Eq. \eqref{C_n}}. In our plots, we specifically set $\left(\lambda/2\pi H_0^2\right)^{1/4}= 10$. We remark that the growth and oscillating frequency differ from the quadratic case. However, just as in the quadratic scenario, the background $\bar\varphi$ and the fluctuation $\delta\varphi$ share the same oscillating frequency.} 

\subsection{Comparison between models}

We reproduced the dustlike behavior of the background quadratic potential (Eq.~\eqref{density_pressure_p2}), as well as the radiationlike behavior of the quartic potential in Eq.~\eqref{density_pressure_p4}. These are well known properties of the cosmological evolution of homogeneous scalar fields (zeroth order in the gradient expansion), as the solutions in Eqs.~\eqref{delta_dust} and \eqref{delta_rad} show, (see also Fig. \ref{fig1}). Additionally, the  growth rate of inhomogeneities in the superhorizon scale is faster for the density contrast in the quartic case ($\propto t$) than for the quadratic potential ($\propto t^{2/3}$). 

We must emphasize that the solutions of this section show that the matter fields present the same time dependence as their equivalents from linear cosmological perturbation theory. In fact the continuity and Euler equations (Eq.~\eqref{density_ev}) are equivalent to Eqs. (8.32) and (8.33) in Ref. \cite{Malik}, expressed for an arbitrary gauge. On the other hand, the dominant contribution of the metric fluctuations at superhorizon scales comes from the zeroth order, time-independent profiles of $\psi$, expressed in Eq.~\eqref{barpsi}. This term is responsible for the conservation of the curvature perturbation on superhorizon scales at all orders in a perturbative expansion in a (multiple) fluid-dominated universe (see e.g.~\cite{Wands:2000dp,Lyth:2004gb,langlois2005evolution,bruni2014einstein}), and in a scalar-field-dominated universe (see e.g.~\citep{Starobinsky:1985ibc,unruh1998cosmological,liddle2000super,Malik:2005cy}). 
The zeroth order metric fluctuation is the  source of the second-order matter fields since they are related through the relativistic version of the Poisson equation (see, e.g., \cite{peebles2020large,wands-slosar,Hidalgo:2013mba}).

\section{Discussion}\label{conclusions}

After showing in Sec. \ref{section2} how the energy-momentum tensor of a CSF can be expressed in terms of perfect fluid variables, in Sec. \ref{section3} we presented the EKG system of equations in a $3+1$ formalism. Also, in Sec. \ref{section3} we showed how to rewrite the system in the cosmological scenario, by reexpressing it in terms of a (cosmological) conformal decomposition. In Sec. \ref{section4} we presented the gradient expansion, employed to obtain the system of equations and the solutions valid to zeroth and second order in the ratio $H^{-1}/L$. The solutions derived here may be used to describe a universe dominated by a CSF at the background with superhorizon inhomogeneities. Finally, in Sec. \ref{section5} we applied our results to two simple examples, namely a quadratic and a quartic potential. They both have been proved useful in the description of dark matter models and reheating scenarios. At the background level, we were able to reproduce the results previously obtained in the literature. We also found the solutions of the inhomogeneous variables of the system for each of these potentials, which have not been previously reported in the literature. 

The perfect fluid description of the complex scalar field is a common practice \citep{complexsf1,complexsf3,charge4,chavanismipaper,jeans1,jeans2, charge1,li2014cosmological,RS,shapiro02, shapiro03, mipaper,Padilla:2020jdj}. However, the use of this description is valid as long as there are no nodes in the distribution of the CSF inomogeneities. Typically, these nodes are expected during the structure formation process, at highly nonlinear stages. The regime of the solutions we obtained in this article pertains much larger scales, and, therefore, the solutions remain valid.

  As mentioned earlier, the solutions here provided are valid for cosmological fluctuations much larger in size than the cosmological horizon. Moreover, the gradient expansion formalism we used has the advantage of not imposing any restriction on the amplitude of the inhomogeneities, as opposed to the standard, linear theory of cosmological perturbations, which is valid only for small amplitudes. In this way, our description allows the study of inhomogeneities of any amplitude, ideal to assess the formation of PBHs. {Our solutions are useful as initial conditions of numerical codes solving the EKG system which encodes all the relativistic effects at play}. This is crucial in the accurate study of the origin supermassive and/or primordial black holes.

The above results bring important consequences for the evolution of fluctuations in a universe dominated by a canonical CSF. In the reheating scenario, setting initial conditions at the inflationary stage, fluctuations may reach an amplitude at the horizon-crossing time, dependent on the potential  of the dominating field at the fast-oscillations period. Typically, a critical amplitude of fluctuations at horizon crossing is defined as a criterion to reach either the formation of structures (for amplitudes above the threshold) or a dilution of fluctuations. The difference in the evolution of matter fluctuations at superhorizon scales implies that the critical amplitude for collapse must strongly depend on the model. 

The existence of a critical amplitude can be inferred from the fact that a scalar field modeling dark matter usually shows characteristic dilution scales (the instability scale for each particular model). 
In the case of an oscillating scalar field the numerical values for the threshold amplitude are, to the best of our knowledge, still to be determined  (for perturbative analytical approximations to the matter density threshold amplitude, in the RSF scenario, see e.g.~\cite{hidalgo,auclair2020primordial,martin_2019}, and for a numerical study see \cite{Torres-Lomas:2014bua}). We shall explore this aspect for the complex scalar field in a follow-up study.


\acknowledgments
The authors are grateful to Tonatiuh Matos and Olivier Sarbach for useful discussions. This work is sponsored by CONACyT Network Projects No. 376127 `Sombras, lentes y ondas gravitatorias generadas por objetos compactos astrof\'isicos', No. 304001 `Estudio de campos escalares con aplicaciones en cosmolog\'ia y astrof\'isica'. LEP and JCH acknowledge sponsorship from CONACyT through grant CB-2016-282569 and from Program UNAM-PAPIIT Grant IN107521 "Sector Oscuro y Agujeros Negros Primordiales". 
\appendix 

\section{The 4-current conservation equation}\label{Ap:current}

The 4-current conservation equation \eqref{4current} can be re-expressed as
\begin{equation}
    \nabla_\mu \mathcal{J}^\mu = \frac{1}{\sqrt{-g}}\partial_\mu(\sqrt{-g}\mathcal{J}^\mu), \ \ \ \ \ \text{where} \ \ \ \ \ \mathcal{J}^\mu = -i\left[\varphi^*\nabla^\nu\varphi-\varphi\nabla^\nu\varphi^*\right].
\end{equation}
Expanding the above equation, using that $\sqrt{-g}=\alpha\sqrt{\gamma}$, and the conformal cosmological decomposition $\sqrt{\gamma} = \psi^6a^3\sqrt{\eta}$, we have
\begin{eqnarray}\label{4current2}
    && -\frac{1}{\alpha\psi^6a^3\sqrt{\eta}}\left\lbrace\partial_t \left[\alpha \psi^6a^3\sqrt{\eta}\left(-\frac{1}{\alpha^2}(\varphi^*\partial_t\varphi-\varphi\partial_t\varphi^*)+\frac{\beta^i}{\alpha^2}(\varphi^*\partial_i\varphi-\varphi\partial_i\varphi^*)\right)\right]+\partial_i\left[\alpha\psi^6a^3\sqrt{\eta}\left(\frac{\beta^i}{\alpha^2}(\varphi^*\partial_t\varphi-\varphi\partial_t\varphi^*)\right.\right.\right.\nonumber \\
    && \left.\left.\left.+\left(\gamma^{ij}-\frac{\beta^i\beta^j}{\alpha^2}\right)(\varphi^*\partial_j\varphi-\varphi\partial_j\varphi^*)\right)\right]\right\rbrace = 0.
\end{eqnarray}
The $O(\epsilon^0)$ of the above equation results
\begin{equation}\label{currentbg1}
    \partial_t[a^3(\bar \varphi^*\partial_t\bar \varphi-\bar \varphi\partial_t\bar \varphi^*)]=O(\epsilon^2).
\end{equation}
On the other hand, the $O(\epsilon^2)$ of \eqref{4current2} is given by
\begin{equation}
    \partial_t \left\lbrace(6\Psi-\Phi)a^3(\bar \varphi^*\partial_t\bar \varphi-\bar \varphi\partial_t\bar \varphi^*)+a^3\left[((\bar \varphi^*\partial_t\delta\varphi-\bar \varphi\partial_t\delta\varphi^*))+(\delta\varphi^*\partial_t\bar \varphi-\delta\varphi\partial_t\bar \varphi^*)\right]\right\rbrace = O(\epsilon^4).
\end{equation}
Using \eqref{eq57} we have $6\partial_t \Psi = 3H \Phi$, and then $6\partial_t\Psi-\partial_t\Phi =3H\Phi-\partial_t\Phi = -a^3\partial_t(\Phi/a^3)$. Then, we can rewrite the above equation as
\begin{equation}\label{currentapp}
    \partial_t\left\lbrace a^3[\partial_t(\bar\varphi^*\delta\varphi-\bar\varphi\delta\varphi^*)-2(\partial_t\bar\varphi^*\delta\varphi-\partial_t\bar\varphi\delta\varphi^*)]\right\rbrace-a^3\partial_t\left(\frac{\Phi}{a^3}\right)[a^3(\bar\varphi^*\partial_t\bar\varphi-\bar\varphi\partial_t\bar\varphi^*)]=O(\epsilon^4),
\end{equation}
where in the above equation we have used \eqref{currentbg1} to simplify the expression.

\bibliography{references}

\begin{thebibliography}{100}

\bibitem{zee2010quantum}
A.~Zee, {\em Quantum field theory in a nutshell}, vol.~7.
\newblock Princeton university press, 2010.

\bibitem{dodelson2003modern}
S.~Dodelson, {\em Modern cosmology}.
\newblock Elsevier, 2003.

\bibitem{bs1}
D.~J. Kaup, ``{Klein-Gordon Geon},'' {\em Phys. Rev.}, vol.~172,
  pp.~1331--1342, 1968.

\bibitem{bs2}
R.~Ruffini and S.~Bonazzola, ``Systems of self-gravitating particles in general
  relativity and the concept of an equation of state,'' {\em Phys. Rev.},
  vol.~187, pp.~1767--1783, Nov 1969.

\bibitem{bs3}
W.~Thirring, ``Bosonic black holes,'' {\em Physics Letters B}, vol.~127, no.~1,
  pp.~27 -- 29, 1983.

\bibitem{bs4}
J.~Breit, S.~Gupta, and A.~Zaks, ``Cold bose stars,'' {\em Physics Letters B},
  vol.~140, no.~5, pp.~329 -- 332, 1984.

\bibitem{bs5}
E.~Takasugi and M.~Yoshimura, ``{Gravitational Stability of Cold Bose Star},''
  {\em Z. Phys. C}, vol.~26, p.~241, 1984.

\bibitem{bs6}
M.~Colpi, S.~Shapiro, and I.~Wasserman, ``{Boson Stars: Gravitational
  Equilibria of Selfinteracting Scalar Fields},'' {\em Phys. Rev. Lett.},
  vol.~57, pp.~2485--2488, 1986.

\bibitem{bs7}
M.~Gleiser, ``Stability of boson stars,'' {\em Phys. Rev. D}, vol.~38,
  pp.~2376--2385, Oct 1988.

\bibitem{bs9}
R.~Ferrell and M.~Gleiser, ``{Gravitational Atoms. 1. Gravitational Radiation
  From Excited Boson Stars},'' {\em Phys. Rev. D}, vol.~40, p.~2524, 1989.

\bibitem{bs10}
M.~Gleiser and R.~Watkins, ``{Gravitational Stability of Scalar Matter},'' {\em
  Nucl. Phys. B}, vol.~319, pp.~733--746, 1989.

\bibitem{bs11}
E.~Seidel and W.-M. Suen, ``{Dynamical Evolution of Boson Stars. 1. Perturbing
  the Ground State},'' {\em Phys. Rev. D}, vol.~42, pp.~384--403, 1990.

\bibitem{bs12}
P.~Jetzer, ``{Boson stars},'' {\em Phys. Rept.}, vol.~220, pp.~163--227, 1992.

\bibitem{bs13}
E.~Seidel and W.-M. Suen, ``{Formation of solitonic stars through gravitational
  cooling},'' {\em Phys. Rev. Lett.}, vol.~72, pp.~2516--2519, 1994.

\bibitem{bs14}
J.~Balakrishna, E.~Seidel, and W.-M. Suen, ``{Dynamical evolution of boson
  stars. 2. Excited states and selfinteracting fields},'' {\em Phys. Rev. D},
  vol.~58, p.~104004, 1998.

\bibitem{Matos:2000ng}
T.~Matos and L.~Urena-Lopez, ``{Quintessence and scalar dark matter in the
  universe},'' {\em Class. Quant. Grav.}, vol.~17, pp.~L75--L81, 2000.

\bibitem{Matos:1998vk}
T.~Matos and F.~S. Guzman, ``{Scalar fields as dark matter in spiral
  galaxies},'' {\em Class. Quant. Grav.}, vol.~17, pp.~L9--L16, 2000.

\bibitem{Matos:1999et}
T.~Matos, F.~S. Guzman, and L.~Urena-Lopez, ``{Scalar field as dark matter in
  the universe},'' {\em Class. Quant. Grav.}, vol.~17, pp.~1707--1712, 2000.

\bibitem{Peebles2000}
P.~J.~E. {Peebles}, ``{Fluid Dark Matter},'' {\em apjl}, vol.~534,
  pp.~L127--L129, May 2000.

\bibitem{Goodman:2000tg}
J.~Goodman, ``{Repulsive dark matter},'' {\em New Astron.}, vol.~5, p.~103,
  2000.

\bibitem{Sahni:1999qe}
V.~Sahni and L.~Wang, ``New cosmological model of quintessence and dark
  matter,'' {\em Phys. Rev. D}, vol.~62, p.~103517, Oct 2000.

\bibitem{charge2}
A.~Arbey, J.~Lesgourgues, and P.~Salati, ``Cosmological constraints on
  quintessential halos,'' {\em Physical Review D}, vol.~65, no.~8, p.~083514,
  2002.

\bibitem{Arbey:2003sj}
A.~Arbey, J.~Lesgourgues, and P.~Salati, ``{Galactic halos of fluid dark
  matter},'' {\em Phys. Rev. D}, vol.~68, p.~023511, 2003.

\bibitem{Cedeno:2017sou}
F.~X.~L. Cede\~no, A.~X. Gonz\'alez-Morales, and L.~A. Ure\~na L\'opez,
  ``Cosmological signatures of ultralight dark matter with an axionlike
  potential,'' {\em Phys. Rev. D}, vol.~96, p.~061301, Sep 2017.

\bibitem{rev1}
J.~Magana and T.~Matos, ``{A brief Review of the Scalar Field Dark Matter
  model},'' {\em J. Phys. Conf. Ser.}, vol.~378, p.~012012, 2012.

\bibitem{rev2}
A.~Su\'arez, V.~H. Robles, and T.~Matos, ``{A Review on the Scalar
  Field/Bose-Einstein Condensate Dark Matter Model},'' {\em Astrophys. Space
  Sci. Proc.}, vol.~38, pp.~107--142, 2014.

\bibitem{Marsh:2015xka}
D.~J.~E. Marsh, ``{Axion Cosmology},'' {\em Phys. Rept.}, vol.~643, pp.~1--79,
  2016.

\bibitem{rev4}
L.~Hui, J.~P. Ostriker, S.~Tremaine, and E.~Witten, ``{Ultralight scalars as
  cosmological dark matter},'' {\em Phys. Rev. D}, vol.~95, no.~4, p.~043541,
  2017.

\bibitem{niemeyer2019small}
J.~C. Niemeyer, ``Small-scale structure of fuzzy and axion-like dark matter,''
  {\em Progress in Particle and Nuclear Physics}, vol.~113, p.~103787, 2020.

\bibitem{RS}
T.~Rindler-Daller and P.~R. Shapiro, ``{Complex scalar field dark matter on
  galactic scales},'' {\em Mod. Phys. Lett. A}, vol.~29, no.~2, p.~1430002,
  2014.

\bibitem{ferreira2020ultra}
E.~G. Ferreira, ``Ultra-light dark matter,'' {\em arXiv preprint
  arXiv:2005.03254}, 2020.

\bibitem{de1}
P.~Peebles and B.~Ratra, ``{The Cosmological Constant and Dark Energy},'' {\em
  Rev. Mod. Phys.}, vol.~75, pp.~559--606, 2003.

\bibitem{de2}
S.~M. Carroll, ``{The Cosmological constant},'' {\em Living Rev. Rel.}, vol.~4,
  p.~1, 2001.

\bibitem{de3}
R.~Rakhi and K.~Indulekha, ``{Dark Energy and Tracker Solution: A Review},'' 10
  2009.

\bibitem{quintessence}
S.~Tsujikawa, ``{Quintessence: A Review},'' {\em Class. Quant. Grav.}, vol.~30,
  p.~214003, 2013.

\bibitem{Zlatev:1998tr}
I.~Zlatev, L.-M. Wang, and P.~J. Steinhardt, ``{Quintessence, cosmic
  coincidence, and the cosmological constant},'' {\em Phys. Rev. Lett.},
  vol.~82, pp.~896--899, 1999.

\bibitem{Corasaniti:2002vg}
P.~S. Corasaniti and E.~Copeland, ``{A Model independent approach to the dark
  energy equation of state},'' {\em Phys. Rev. D}, vol.~67, p.~063521, 2003.

\bibitem{de4}
R.~Caldwell, R.~Dave, and P.~J. Steinhardt, ``{Cosmological imprint of an
  energy component with general equation of state},'' {\em Phys. Rev. Lett.},
  vol.~80, pp.~1582--1585, 1998.

\bibitem{de5}
R.~R. {Caldwell}, ``{A phantom menace? Cosmological consequences of a dark
  energy component with super-negative equation of state},'' {\em Physics
  Letters B}, vol.~545, pp.~23--29, Oct. 2002.

\bibitem{de6}
R.~Caldwell, M.~Kamionkowski, and N.~Weinberg, ``Phantom energy: Dark energy
  with $w < - 1$ causes a cosmic doomsday,'' {\em Physical review letters},
  vol.~91, p.~071301, 09 2003.

\bibitem{de7}
M.~P. Dabrowski, C.~Kiefer, and B.~Sandhofer, ``{Quantum phantom cosmology},''
  {\em Phys. Rev. D}, vol.~74, p.~044022, 2006.

\bibitem{de8}
S.~Capozziello, S.~Nojiri, and S.~Odintsov, ``{Unified phantom cosmology:
  Inflation, dark energy and dark matter under the same standard},'' {\em Phys.
  Lett. B}, vol.~632, pp.~597--604, 2006.

\bibitem{de9}
A.~Sen, ``{Rolling tachyon},'' {\em JHEP}, vol.~04, p.~048, 2002.

\bibitem{de10}
S.~M. Carroll, ``Quintessence and the rest of the world: Suppressing long-range
  interactions,'' {\em Phys. Rev. Lett.}, vol.~81, pp.~3067--3070, Oct 1998.

\bibitem{de11}
L.~Amendola and S.~Tsujikawa, {\em {Dark Energy}: {Theory and Observations}}.
\newblock Cambridge University Press, 1 2015.

\bibitem{Linde1982}
A.~D. {Linde}, ``{A new inflationary universe scenario: A possible solution of
  the horizon, flatness, homogeneity, isotropy and primordial monopole
  problems},'' {\em Physics Letters B}, vol.~108, pp.~389--393, Feb. 1982.

\bibitem{inf3}
A.~D. Linde, {\em {Particle physics and inflationary cosmology}}, vol.~5.
\newblock 1990.

\bibitem{inf4}
K.~Olive, ``Inflation,'' {\em Physics Reports}, vol.~190, pp.~307--403, June
  1990.

\bibitem{Guth}
A.~H. Guth, ``Inflationary universe: A possible solution to the horizon and
  flatness problems,'' {\em Physical Review D}, vol.~23, no.~2, p.~347, 1981.

\bibitem{Lucchin:1984yf}
F.~Lucchin and S.~Matarrese, ``{Power Law Inflation},'' {\em Phys. Rev. D},
  vol.~32, p.~1316, 1985.

\bibitem{Liddle:1999mq}
A.~R. Liddle, ``{An Introduction to cosmological inflation},'' in {\em {ICTP
  Summer School in High-Energy Physics and Cosmology}}, pp.~260--295, 1 1999.

\bibitem{Ratra:1989uz}
B.~Ratra, ``{Inflation in an Exponential Potential Scalar Field Model},'' {\em
  Phys. Rev. D}, vol.~45, pp.~1913--1952, 1992.

\bibitem{DiMarco:2018bnw}
A.~Di~Marco, G.~Pradisi, and P.~Cabella, ``{Inflationary scale, reheating
  scale, and pre-BBN cosmology with scalar fields},'' {\em Phys. Rev. D},
  vol.~98, no.~12, p.~123511, 2018.

\bibitem{inf2}
A.~Albrecht, P.~J. Steinhardt, M.~S. Turner, and F.~Wilczek, ``{Reheating an
  Inflationary Universe},'' {\em Phys. Rev. Lett.}, vol.~48, p.~1437, 1982.

\bibitem{Vazquez:2018qdg}
J.~A. V\'azquez, L.~E. Padilla, and T.~Matos, ``{Inflationary Cosmology: From
  Theory to Observations},'' 10 2018.

\bibitem{16}
T.~Matos and L.~A. Urena-Lopez, ``Further analysis of a cosmological model with
  quintessence and scalar dark matter,'' {\em Physical Review D}, vol.~63,
  no.~6, p.~063506, 2001.

\bibitem{9}
W.~Hu, R.~Barkana, and A.~Gruzinov, ``Fuzzy cold dark matter: the wave
  properties of ultralight particles,'' {\em Physical Review Letters}, vol.~85,
  no.~6, p.~1158, 2000.

\bibitem{18}
A.~Su{\'a}rez and T.~Matos, ``Structure formation with scalar-field dark
  matter: the fluid approach,'' {\em Monthly Notices of the Royal Astronomical
  Society}, vol.~416, no.~1, pp.~87--93, 2011.

\bibitem{matos2007flat}
T.~Matos and L.~A. Ure{\~n}a-L{\'o}pez, ``Flat rotation curves in scalar field
  galaxy halos,'' {\em General Relativity and Gravitation}, vol.~39, no.~8,
  pp.~1279--1286, 2007.

\bibitem{robles2012flat}
V.~H. Robles and T.~Matos, ``Flat central density profile and constant dark
  matter surface density in galaxies from scalar field dark matter,'' {\em
  Monthly Notices of the Royal Astronomical Society}, vol.~422, no.~1,
  pp.~282--289, 2012.

\bibitem{li2014cosmological}
B.~Li, T.~Rindler-Daller, and P.~R. Shapiro, ``Cosmological constraints on
  bose-einstein-condensed scalar field dark matter,'' {\em Physical Review D},
  vol.~89, no.~8, p.~083536, 2014.

\bibitem{Carrion:2021yeh}
K.~Carrion, J.~C. Hidalgo, A.~Montiel, and L.~E. Padilla, ``{Complex Scalar
  Field Reheating and Primordial Black Hole production},'' 1 2021.

\bibitem{reh1}
B.~A. Bassett, S.~Tsujikawa, and D.~Wands, ``{Inflation dynamics and
  reheating},'' {\em Rev. Mod. Phys.}, vol.~78, pp.~537--589, 2006.

\bibitem{reh2}
R.~Allahverdi, R.~Brandenberger, F.-Y. Cyr-Racine, and A.~Mazumdar,
  ``{Reheating in Inflationary Cosmology: Theory and Applications},'' {\em Ann.
  Rev. Nucl. Part. Sci.}, vol.~60, pp.~27--51, 2010.

\bibitem{reh3}
A.~V. Frolov, ``{Non-linear Dynamics and Primordial Curvature Perturbations
  from Preheating},'' {\em Class. Quant. Grav.}, vol.~27, p.~124006, 2010.

\bibitem{reh4}
M.~A. Amin, M.~P. Hertzberg, D.~I. Kaiser, and J.~Karouby, ``{Nonperturbative
  Dynamics Of Reheating After Inflation: A Review},'' {\em Int. J. Mod. Phys.
  D}, vol.~24, p.~1530003, 2014.

\bibitem{Kofman:1997yn}
L.~Kofman, A.~D. Linde, and A.~A. Starobinsky, ``{Towards the theory of
  reheating after inflation},'' {\em Phys. Rev. D}, vol.~56, pp.~3258--3295,
  1997.

\bibitem{sfdmrh1}
N.~Musoke, S.~Hotchkiss, and R.~Easther, ``{Lighting the Dark: Evolution of the
  Postinflationary Universe},'' {\em Phys. Rev. Lett.}, vol.~124, no.~6,
  p.~061301, 2020.

\bibitem{sfdmrh2}
J.~C. Niemeyer and R.~Easther, ``{Inflaton clusters and inflaton stars},'' {\em
  JCAP}, vol.~07, p.~030, 2020.

\bibitem{Eggemeier:2020zeg}
B.~Eggemeier, J.~C. Niemeyer, and R.~Easther, ``{Formation of Inflaton Halos
  after Inflation},'' 11 2020.

\bibitem{Malik:2006ir}
K.~A. Malik, ``{A not so short note on the Klein-Gordon equation at second
  order},'' {\em JCAP}, vol.~03, p.~004, 2007.

\bibitem{compinf1}
A.~Kamenshchik, I.~Khalatnikov, and A.~Toporensky, ``{Complex inflaton field in
  quantum cosmology},'' {\em Int. J. Mod. Phys. D}, vol.~6, pp.~649--672, 1997.

\bibitem{compinf2}
I.~Khalatnikov and A.~Mezhlumian, ``{The Classical and quantum cosmology with a
  complex scalar field},'' {\em Phys. Lett. A}, vol.~169, pp.~308--312, 1992.

\bibitem{compinf3}
I.~Khalatnikov and P.~Schiller, ``{From instanton to inflationary universe},''
  {\em Phys. Lett. B}, vol.~302, pp.~176--182, 1993.

\bibitem{compinf4}
L.~Amendola, I.~M. Khalatnikov, M.~Litterio, and F.~Occhionero, ``Quantum
  cosmology with a complex field,'' {\em Phys. Rev. D}, vol.~49,
  pp.~1881--1885, Feb 1994.

\bibitem{compinf5}
R.~H.~S. Budhi, S.~Kashiwase, and D.~Suematsu, ``{Inflation in a modified
  radiative seesaw model},'' {\em Phys. Rev. D}, vol.~90, no.~11, p.~113013,
  2014.

\bibitem{compinf6}
R.~H.~S. Budhi, S.~Kashiwase, and D.~Suematsu, ``Constrained inflaton due to a
  complex scalar,'' {\em Journal of Cosmology and Astroparticle Physics},
  vol.~2015, pp.~039--039, sep 2015.

\bibitem{compinf7}
K.~Kannike, A.~Kubarski, L.~Marzola, and A.~Racioppi, ``{A minimal model of
  inflation and dark radiation},'' {\em Phys. Lett. B}, vol.~792, pp.~74--80,
  2019.

\bibitem{compinf8}
G.~Barenboim and W.-I. Park, ``{Spiral Inflation},'' {\em Phys. Lett. B},
  vol.~741, pp.~252--255, 2015.

\bibitem{compinf9}
G.~Barenboim and W.-I. Park, ``{Spiral Inflation with Coleman-Weinberg
  Potential},'' {\em Phys. Rev. D}, vol.~91, no.~6, p.~063511, 2015.

\bibitem{compinf10}
J.~McDonald, ``{Sub-Planckian Two-Field Inflation Consistent with the Lyth
  Bound},'' {\em JCAP}, vol.~09, p.~027, 2014.

\bibitem{compinf11}
G.~Barenboim and W.-I. Park, ``{Spontaneous baryogenesis in spiral
  inflation},'' {\em Eur. Phys. J. C}, vol.~79, no.~6, p.~456, 2019.

\bibitem{salopek1990nonlinear}
D.~Salopek and J.~Bond, ``Nonlinear evolution of long-wavelength metric
  fluctuations in inflationary models,'' {\em Physical Review D}, vol.~42,
  no.~12, p.~3936, 1990.

\bibitem{tanaka2007gradient}
Y.~Tanaka and M.~Sasaki, ``Gradient expansion approach to nonlinear
  superhorizon perturbations. ii: —a single scalar field—,'' {\em Progress
  of Theoretical Physics}, vol.~118, no.~3, pp.~455--473, 2007.

\bibitem{kodama1998evolution}
H.~Kodama and T.~Hamazaki, ``Evolution of cosmological perturbations in the
  long wavelength limit,'' {\em Physical Review D}, vol.~57, no.~12, p.~7177,
  1998.

\bibitem{Takamizu:2018uty}
Y.-i. Takamizu, ``{Gradient expansion formalism for nonlinear superhorizon
  perturbations},'' 4 2018.

\bibitem{Sasaki:1998ug}
M.~Sasaki and T.~Tanaka, ``{Superhorizon scale dynamics of multiscalar
  inflation},'' {\em Prog. Theor. Phys.}, vol.~99, pp.~763--782, 1998.

\bibitem{Malik:2005cy}
K.~A. Malik, ``{Gauge-invariant perturbations at second order: Multiple scalar
  fields on large scales},'' {\em JCAP}, vol.~11, p.~005, 2005.

\bibitem{Babichev:2018twg}
E.~Babichev, S.~Ramazanov, and A.~Vikman, ``{Recovering $P(X)$ from a canonical
  complex field},'' {\em JCAP}, vol.~11, p.~023, 2018.

\bibitem{shibata1999black}
M.~Shibata and M.~Sasaki, ``Black hole formation in the friedmann universe:
  Formulation and computation in numerical relativity,'' {\em Physical Review
  D}, vol.~60, no.~8, p.~084002, 1999.

\bibitem{harada2015cosmological}
T.~Harada, C.-M. Yoo, T.~Nakama, and Y.~Koga, ``Cosmological long-wavelength
  solutions and primordial black hole formation,'' {\em Physical Review D},
  vol.~91, no.~8, p.~084057, 2015.

\bibitem{tanaka2007gradient2}
Y.~Tanaka and M.~Sasaki, ``Gradient expansion approach to nonlinear
  superhorizon perturbations,'' {\em Progress of Theoretical Physics},
  vol.~117, no.~4, pp.~633--654, 2007.

\bibitem{madelung1927quantentheorie}
E.~Madelung, ``Quantentheorie in hydrodynamischer form,'' {\em Zeitschrift
  f{\"u}r Physik}, vol.~40, no.~3-4, pp.~322--326, 1927.

\bibitem{charge4}
A.~Su{\'a}rez and P.-H. Chavanis, ``Cosmological evolution of a complex scalar
  field with repulsive or attractive self-interaction,'' {\em Physical Review
  D}, vol.~95, no.~6, p.~063515, 2017.

\bibitem{complexsf4}
B.~Li, T.~Rindler-Daller, and P.~R. Shapiro, ``{Cosmological Constraints on
  Bose-Einstein-Condensed Scalar Field Dark Matter},'' {\em Phys. Rev. D},
  vol.~89, no.~8, p.~083536, 2014.

\bibitem{Carvente:2020aae}
B.~Carvente, V.~Jaramillo, C.~Escamilla-Rivera, and D.~N\'u\~nez,
  ``{Observational constraints on complex quintessence with attractive
  self-interaction},'' 8 2020.

\bibitem{jeans2}
A.~Su{\'a}rez and P.-H. Chavanis, ``Hydrodynamic representation of the
  klein-gordon-einstein equations in the weak field limit,'' in {\em Journal of
  Physics: Conference Series}, vol.~654, p.~012008, IOP Publishing, 2015.

\bibitem{jeans3}
A.~Su{\'a}rez and P.-H. Chavanis, ``Hydrodynamic representation of the
  klein-gordon-einstein equations in the weak field limit: General formalism
  and perturbations analysis,'' {\em Physical Review D}, vol.~92, no.~2,
  p.~023510, 2015.

\bibitem{mipaper}
L.~E. Padilla, T.~Rindler-Daller, P.~R. Shapiro, T.~Matos, and J.~A. V\'azquez,
  ``{On the Core-Halo Mass Relation in Scalar Field Dark Matter Models and its
  Consequences for the Formation of Supermassive Black Holes},'' {\em Accepted
  to be published at PRD}, 10 2020.

\bibitem{chavanismipaper}
P.-H. Chavanis, ``Predictive model of bec dark matter halos with a solitonic
  core and an isothermal atmosphere,'' {\em Physical Review D}, vol.~100,
  no.~8, p.~083022, 2019.

\bibitem{shapiro03}
T.~Rindler-Daller and P.~R. Shapiro, ``Angular momentum and vortex formation in
  bose--einstein-condensed cold dark matter haloes,'' {\em Monthly Notices of
  the Royal Astronomical Society}, vol.~422, no.~1, pp.~135--161, 2012.

\bibitem{Alcubierre_adm}
J.~M. Torres, M.~Alcubierre, A.~Diez-Tejedor, and D.~N\'u\~nez, ``Cosmological
  nonlinear structure formation in full general relativity,'' {\em Phys. Rev.
  D}, vol.~90, p.~123002, Dec 2014.

\bibitem{Alcubierre:2005gh}
M.~Alcubierre, A.~Corichi, J.~A. Gonzalez, D.~Nunez, B.~Reimann, and
  M.~Salgado, ``{Generalized harmonic spatial coordinates and hyperbolic shift
  conditions},'' {\em Phys. Rev. D}, vol.~72, p.~124018, 2005.

\bibitem{charge1}
A.~Su{\'a}rez and P.-H. Chavanis, ``Hydrodynamic representation of the
  klein-gordon-einstein equations in the weak field limit: General formalism
  and perturbations analysis,'' {\em Physical Review D}, vol.~92, no.~2,
  p.~023510, 2015.

\bibitem{charge3}
J.-A. Gu and W.-Y. Hwang, ``Can the quintessence be a complex scalar field?,''
  {\em Physics Letters B}, vol.~517, no.~1-2, pp.~1--6, 2001.

\bibitem{piran1985inflation}
T.~Piran and R.~M. Williams, ``Inflation in universes with a massive scalar
  field,'' {\em Physics Letters B}, vol.~163, no.~5-6, pp.~331--335, 1985.

\bibitem{Padilla:2019fju}
L.~E. Padilla, J.~A. V\'azquez, T.~Matos, and G.~Germ\'an, ``{Scalar Field Dark
  Matter Spectator During Inflation: The Effect of Self-interaction},'' {\em
  JCAP}, vol.~05, p.~056, 2019.

\bibitem{Sahni:2001qp}
V.~Sahni, M.~Sami, and T.~Souradeep, ``{Relic gravity waves from brane world
  inflation},'' {\em Phys. Rev. D}, vol.~65, p.~023518, 2002.

\bibitem{Hidalgo:2011fj}
J.~Hidalgo, L.~Urena-Lopez, and A.~R. Liddle, ``{Unification models with
  reheating via Primordial Black Holes},'' {\em Phys. Rev. D}, vol.~85,
  p.~044055, 2012.

\bibitem{Jetzer}
P.~Jetzer and D.~Scialom, ``{Time evolution of the perturbations for a complex
  scalar field in Friedmann-Lemaitre universe},'' {\em Phys. Rev.}, vol.~D55,
  pp.~7440--7450, 1997.

\bibitem{Malik}
K.~A. Malik and D.~Wands, ``{Cosmological perturbations},'' {\em Phys. Rept.},
  vol.~475, pp.~1--51, 2009.

\bibitem{Wands:2000dp}
D.~Wands, K.~A. Malik, D.~H. Lyth, and A.~R. Liddle, ``{A New approach to the
  evolution of cosmological perturbations on large scales},'' {\em Phys. Rev.
  D}, vol.~62, p.~043527, 2000.

\bibitem{Lyth:2004gb}
D.~H. Lyth, K.~A. Malik, and M.~Sasaki, ``{A General proof of the conservation
  of the curvature perturbation},'' {\em JCAP}, vol.~05, p.~004, 2005.

\bibitem{langlois2005evolution}
D.~Langlois and F.~Vernizzi, ``Evolution of nonlinear cosmological
  perturbations,'' {\em Physical review letters}, vol.~95, no.~9, p.~091303,
  2005.

\bibitem{bruni2014einstein}
M.~Bruni, J.~C. Hidalgo, and D.~Wands, ``Einstein's signature in cosmological
  large-scale structure,'' {\em The Astrophysical Journal Letters}, vol.~794,
  no.~1, p.~L11, 2014.

\bibitem{Starobinsky:1985ibc}
A.~A. Starobinsky, ``{Multicomponent de Sitter (Inflationary) Stages and the
  Generation of Perturbations},'' {\em JETP Lett.}, vol.~42, pp.~152--155,
  1985.

\bibitem{unruh1998cosmological}
W.~Unruh, ``Cosmological long wavelength perturbations,'' {\em arXiv preprint
  astro-ph/9802323}, 1998.

\bibitem{liddle2000super}
A.~R. Liddle, D.~H. Lyth, K.~A. Malik, and D.~Wands, ``Super-horizon
  perturbations and preheating,'' {\em Physical Review D}, vol.~61, no.~10,
  p.~103509, 2000.

\bibitem{peebles2020large}
P.~J.~E. Peebles, {\em The large-scale structure of the universe}, vol.~98.
\newblock Princeton university press, 2020.

\bibitem{wands-slosar}
D.~Wands and A.~Slosar, ``{Scale-dependent bias from primordial non-Gaussianity
  in general relativity},'' {\em Phys. Rev. D}, vol.~79, p.~123507, 2009.

\bibitem{Hidalgo:2013mba}
J.~C. Hidalgo, A.~J. Christopherson, and K.~A. Malik, ``{The Poisson equation
  at second order in relativistic cosmology},'' {\em JCAP}, vol.~08, p.~026,
  2013.

\bibitem{complexsf1}
P.-H. Chavanis, ``Growth of perturbations in an expanding universe with
  bose-einstein condensate dark matter,'' {\em Astronomy \& Astrophysics},
  vol.~537, p.~A127, 2012.

\bibitem{complexsf3}
P.-H. Chavanis, ``Cosmology with a stiff matter era,'' {\em Physical Review D},
  vol.~92, no.~10, p.~103004, 2015.

\bibitem{jeans1}
A.~Su{\'a}rez and P.-H. Chavanis, ``Jeans-type instability of a complex
  self-interacting scalar field in general relativity,'' {\em Physical Review
  D}, vol.~98, no.~8, p.~083529, 2018.

\bibitem{shapiro02}
B.~Li, P.~R. Shapiro, and T.~Rindler-Daller, ``Bose-einstein-condensed scalar
  field dark matter and the gravitational wave background from inflation: new
  cosmological constraints and its detectability by ligo,'' {\em Physical
  Review D}, vol.~96, no.~6, p.~063505, 2017.

\bibitem{Padilla:2020jdj}
L.~E. Padilla, J.~Sol\'\i{}s-L\'opez, T.~Matos, and A.~\'Avilez-L\'opez,
  ``{Consequences for the Scalar Field Dark Matter model from The McGaugh
  Observed-Baryon Acceleration Correlation},'' 8 2020.

\bibitem{hidalgo}
J.~C. Hidalgo, J.~De~Santiago, G.~German, N.~Barbosa-Cendejas, and
  W.~Ruiz-Luna, ``{Collapse threshold for a cosmological Klein Gordon field},''
  {\em Phys. Rev.}, vol.~D96, no.~6, p.~063504, 2017.

\bibitem{auclair2020primordial}
P.~Auclair and V.~Vennin, ``{Primordial black holes from metric preheating:
  mass fraction in the excursion-set approach},'' 11 2020.

\bibitem{martin_2019}
J.~Martin, T.~Papanikolaou, and V.~Vennin, ``{Primordial black holes from the
  preheating instability in single-field inflation},'' {\em JCAP}, vol.~01,
  p.~024, 2020.

\bibitem{Torres-Lomas:2014bua}
E.~Torres-Lomas, J.~C. Hidalgo, K.~A. Malik, and L.~A. Ure\~na L\'opez,
  ``{Formation of subhorizon black holes from preheating},'' {\em Phys. Rev.
  D}, vol.~89, no.~8, p.~083008, 2014.

\end{thebibliography}
\bibliographystyle{ieeetr}

\end{document}